\documentclass[11pt]{article}

\usepackage{PRIMEarxiv}
\usepackage[utf8]{inputenc} 
\usepackage[T1]{fontenc}    
\usepackage{hyperref}       
\usepackage{booktabs}       
\usepackage{amsfonts}       
\usepackage{nicefrac}       
\usepackage{microtype}      
\usepackage{lipsum}
\usepackage{fancyhdr}       
\usepackage{graphicx}       
\graphicspath{{media/}}     
\usepackage{wrapfig}
\usepackage{times}  
\usepackage{helvet}  
\usepackage{courier}  
\usepackage{natbib}  
\usepackage{caption} 
\usepackage{algorithm}

\usepackage{times}
\usepackage{soul}
\usepackage{url}
\usepackage{amsmath}
\usepackage{multirow}
\usepackage{amsmath,amssymb}
\usepackage{bbm}
\usepackage[noend]{algpseudocode}
\usepackage{mathtools}
\usepackage{mathrsfs}
\usepackage{subcaption}
\usepackage{comment}
\usepackage{outlines}

\newtheorem{definition}{Definition}
\title{Learning to Advise Humans in High-Stakes Settings}

\author{
  Nicholas Wolczynski\\
  University of Texas at Austin \\
  \texttt{nicholas@mccombs.utexas.edu} \\
   \And
  Maytal Saar-Tsechansky \thanks{Equal contribution} \\
   University of Texas at Austin \\
   \texttt{maytal@mail.utexas.edu} \\
   \AND
   Tong Wang  $^*$ \\
  University of Iowa \\
  \texttt{tong-wang@uiowa.edu} \\
}

\begin{document}

\maketitle

\begin{abstract}
Expert decision-makers (DMs) in high-stakes AI-assisted decision-making (AIaDM) settings receive and reconcile recommendations from AI systems before making their final decisions. 
We identify distinct properties of these settings which are key to developing AIaDM models that effectively benefit team performance. First, DMs incur \emph{reconciliation costs} from exerting decision-making resources (e.g., time and effort) when reconciling AI recommendations that contradict their own judgment. Second, DMs in AIaDM settings exhibit \emph{algorithm discretion behavior} (ADB), i.e., an idiosyncratic tendency to imperfectly accept or reject algorithmic recommendations for any given decision task. The human's reconciliation costs and imperfect discretion behavior introduce the need to develop AI systems which (1) provide recommendations selectively, (2) leverage the human partner's ADB to maximize the team's decision accuracy while regularizing for reconciliation costs, and (3) are inherently interpretable. We refer to the task of developing AI to advise humans in AIaDM settings as \emph{learning to advise} and we address this task by first introducing the AI-assisted Team (AIaT)-Learning Framework. We instantiate our framework to develop TeamRules (\textsc{tr}): an algorithm that produces rule-based models and recommendations for AIaDM settings. \textsc{tr} is optimized to selectively advise a human and to trade-off \emph{reconciliation costs} and \emph{team accuracy} for a given environment by leveraging the human partner's ADB. Evaluations on synthetic and real-world benchmark datasets with a variety of simulated human accuracy and discretion behaviors show that \textsc{tr} robustly improves the team's objective across settings over interpretable, rule-based alternatives. 
\end{abstract}

\section{Introduction}
Advances in machine learning performance and interpretability across  domains have brought about a growing focus on human-AI (HAI) systems to enhance human decision-making ~\citep{cai_human-centered_2019,soares_fair-by-design_2019,green_algorithmic_2021,basu_human_2021,lebovitz_engage_2022}.
Most prior work that developed AI methods for human-AI teams focused on settings where the AI can either make all decisions autonomously or can decide to defer to the human for some tasks \citep{madras_predict_2018, gao_2021, defer_multiple_2021}. In this work, we consider the task of \emph{learning to advise} in \emph{high-stakes} AI-assisted decision making (AIaDM) settings where the human must act as the \emph{final decision-maker} (DM) for all instances. In such settings, the AI does not undertake any decisions autonomously; rather, the AI may only advise the human on some or all instances. The task of learning to advise has proved challenging in practice, and existing AI systems often do not significantly improve teams' final decisions in high-stakes AIaDM settings \citep{green_disparate_2019, lebovitz_engage_2022, green_algorithmic_2021}.

In this work, we first outline key properties of AIaDM contexts that pose challenges to team performance and then propose an AIaDM method that addresses them. In particular, prior work that closely studied experts advised by AI in high-stakes settings highlights that experts incur costs from exerting time and effort to reconcile AI recommendations that contradict their own initial judgment \citep{lebovitz_engage_2022}. Consequently, DMs in AIaDM settings incur costs when the AI system \emph{takes action} and offers recommendations that contradict the DMs' initial judgments. 
Each DM in a given context has a tolerance for additional reconciliation costs, and when the effort required to reconcile contradicting advice is excessive given their tolerance, DMs may ignore beneficial advice or disengage with the AI entirely \citep{lebovitz_engage_2022}. Second, DMs exhibit imperfect \emph{algorithm discretion behavior} (ADB), i.e., an idiosyncratic tendency to imperfectly accept or reject algorithmic recommendations for any given decision task. \citet{bansal_is_2021} demonstrate that AI can improve \emph{team} performance in AIaDM settings when the human DM optimally reconciles algorithmic recommendations. However, prior work has established that humans' discretion towards algorithmic recommendations is unlikely to always be optimal, to the detriment of the team \citep{dietvorst_algorithm_2015,green_disparate,green_principles,stop,bansal_does_2021,complete-me}. Thus, how AI can best improve team performance in the presence of sup-optimal human discretion remains an open problem. Finally, high-stakes AIaDM settings also require AI recommendations that are inherently interpretable \citep{vellido_importance_2020,stop}\footnote{In some key high-risk and regulated domains, model interpretability is required by law \citep{doshi-velez_towards_2017,goodman_european_2017}}. Interpretable advice allows experts to reason about and justify the AI-advised decisions that contradict their initial judgment and to edit the patterns underlying the recommendations \citep{rich,BALAGOPAL2021102101}.

To overcome the challenges posed by high-stakes AIaDM properties, AIaDM systems must offer complementary advice that can effectively enhance the team's performance given any human's arbitrary and imperfect ADB, decision-making ability, and tolerance for incurring reconciliation costs, while also producing inherently interpretable advice that experts can establish the reasoning for. We first propose that AIaDM systems make recommendations \emph{selectively}, not only when the AI is likely to be more accurate than the human, but also when the expected benefit to decision-making outweighs the costs incurred from the human's reconciliation of contradictory AI advice. Importantly, the cost-effectiveness of advice is also informed by the human's discretion behavior, namely, the likelihood of the human accepting the advice. Second, we provide theoretical justification for the potential benefits of leveraging the human's ADB, and we empirically show that AIaDM systems can improve team outcomes by simultaneously leveraging the human's ADB, decision history, and tolerance for reconciliation costs within the AI training objective, allowing for direct optimization over the team's \emph{final decision}. 

We propose a framework and an algorithm to address the challenge of \emph{learning to advise} humans in high-stakes settings. Our framework and algorithm produce a personalized and inherently interpretable rule-based model that provides complementary advice to an individual human DM. Specifically, we make the following contributions:

\begin{itemize}
    \item We identify and consider three key properties of high-stakes AIaDM settings where an imperfect human DM may be advised by an AI but always undertakes the final decision. The human DM (1) incurs costs from reconciling contradicting algorithmic recommendations, (2) can exhibit imperfect ADB towards the AI's advice, and (3) requires interpretable explanations to reason about and edit the AI model's recommendations.
    \item We propose that methods for high-stakes AIaDM settings can produce complementary advice beneficial to team performance when they: (1) provide recommendations selectively, (2) are informed by the human DM's algorithm discretion behavior (ADB), decision history, and tolerance for reconciling advice, and (3) are inherently interpretable.  
    \item We develop a framework for learning to advise humans in high-stakes AIaDM settings with the properties above, and, based on our proposed framework, we develop and evaluate the TeamRules (\textsc{tr}) algorithm which generates an inherently interpretable rule-based model designed to complement a human DM by providing selective advice that leverages the human DM's ADB, decision history, and tolerance for reconciling advice. 
\end{itemize}

We evaluate \textsc{tr}'s performance relative to alternative rule-based methods on three real-world datasets and two synthetic datasets, and when paired with different simulated human behaviors. The results show that \textsc{tr} reliably and effectively leverages the human's ADB to selectively provide recommendations that improve the team's objective over alternatives. Additionally, \textsc{tr} can adapt to the partner's tolerance for incurring additional reconciliation costs by trading away benefits to team decision accuracy for lower reconciliation costs incurred. We empirically demonstrate that our method yields superior team performance relative to benchmarks even under imperfect estimates of the human's ADB, showcasing the robustness of our method.  

\section{Related Work}
In this section, we review related work on human-AI teams, algorithm discretion, and rule-based interpretable models.
\subsection{Human-AI Teams}
Existing literature on human-AI (HAI) teams is broad and considers a variety of perspectives and applications. HAI teams are increasingly deployed for decision-making in a multitude of high-stakes settings, including medicine~\citep{cai_human-centered_2019,BALAGOPAL2021102101} and criminal justice~\citep{soares_fair-by-design_2019}. However, AI systems deployed in high-stakes settings often do not lead to complementary team performance that is superior to both the human's and AI's standalone performance \citep{green_disparate,green_principles,lebovitz_engage_2022}.

Expert DMs in high-stakes settings make many complex decisions under time constraints. When an AI system offers the DM advice that contradicts their initial judgement, the DM incurs reconciliation costs due to the additional time and effort required to reason about the contradicting recommendation \citep{lebovitz_engage_2022}. If the AI system over-burdens the DM with excessive reconciliation costs, the DM may begin to disregard the AI advice entirely, gaining no benefit from the AI. 

Prior work on HAI teams demonstrated that directly optimizing the team's objectives is key to producing AI systems that complement human DMs, often by accounting for their decision history \citep{madras_predict_2018,bansal_is_2021}. Specifically, \citet{bansal_is_2021} consider a setting in which a DM is assumed to exhibit optimal algorithm discretion behavior, such that the DM always accepts the AI decision if it will lead to a higher expected team outcome, and rejects otherwise. 
However, humans' discretion of algorithmic recommendations is not always optimal \citep{stop,dietvorst_algorithm_2015,task-dependent-algorithm-aversion}, and, thus, how AI can best improve team performance in the presence of arbitrary and sup-optimal human discretion behavior is an open problem that we aim to address. In this work we propose that AI systems designed for AIaDM settings should effectively complement humans exhibiting  \emph{arbitrary}  ADB; we propose how a human's arbitrary ADB can be brought to bear and demonstrate the subsequent potential benefits to the team's performance.

We note that the AIaDM setting is distinct from the HAI team \emph{deferral} setting in which the AI system can make decisions autonomously, without human involvement, but may defer to the human on some decisions \citep{madras_predict_2018,wang_augmented_2020,wilder_learning_2020,defer_multiple_2021,gao_2021,Bondi_Koster_Sheahan_Chadwick_Bachrach_Cemgil_Paquet_Dvijotham_2022}. Methods developed for the deferral setting do not need to address the human's discretion behavior and costs of reconciling contradictory advice to improve and evaluate the team's performance. 

\subsection{Algorithm Discretion}
The term \emph{algorithm aversion} was proposed to characterize humans' tendency to avoid relying on an algorithmic recommendation in favor of human judgment, even when the algorithm's performance is superior \citep{dietvorst_algorithm_2015,task-dependent-algorithm-aversion}. However, a human's reconciliation of algorithmic recommendation can also exhibit over-reliance on these recommendations \citep{LOGG201990,MAHMUD2022121390}, or be based on an adequate judgment of relevant factors \citep{kim_when_2020,jussupow_why_2020}. Consequently, in this work, we use the term \emph{Algorithm Discretion} to refer to the human's arbitrary and idiosyncratic pattern to accept or reject algorithmic recommendations for any given decision instance. Existing work has demonstrated that algorithm discretion behavior is predictable and is largely a function of human self-confidence in their own decisions \citep{CHONG2022107018,will_you_accept}. Our method models and aims to leverage a human's arbitrary algorithm discretion behavior to improve human-AI team performance. 

\subsection{Rule-Based Models}
Rule-based models are inherently interpretable and easy to understand because they take the form of sparse decision lists, consisting of a series of if... then statements \citep{vc_rules,rudin2019stop,wang2021scalable,wang2018multi}. This model form offers an inherent reason for every prediction-based recommendation \citep{rules_why/15-AOAS848}, and rule-based models are widely recognized as one of the most intuitively understandable models for their transparent inner structures and good model expressivity \citep{rudin2019stop,wang2021scalable}.
In many high-stakes domains, experts require interpretability to vet on the reasoning underlying the model's predictions \citep{rich}. Additionally, AIaDM systems yield more productive outcomes when experts can directly edit the patterns underlying the recommendations, so as to reflect the knowledge that the AI cannot otherwise learn \citep{rich,BALAGOPAL2021102101,edit_machine}. Consequently, in this work, we provide an inherently interpretable, rule-based approach.

The method we introduce in this paper builds on the \textsc{\normalsize{h}\scriptsize{y}\normalsize{rs}} method \citep{wang_gaining_nodate} originally proposed to offer partial interpretability for black-box models. \textsc{\normalsize{h}\scriptsize{y}\normalsize{rs}} is  an extension of the Bayesian Rule Sets (\textsc{brs}) method \citep{wang2017bayesian} for creating rule set classifiers. While these works do not consider the problem of advising a human, we adapted the \textsc{\normalsize{h}\scriptsize{y}\normalsize{rs}} and \textsc{brs} methods as benchmarks to evaluate the benefits of leveraging ADB and considering reconciliation costs within a class of methods.

\section{Leveraging Human Behavior}
\begin{figure}[t]
    \centering
    \includegraphics[width=0.775\columnwidth]{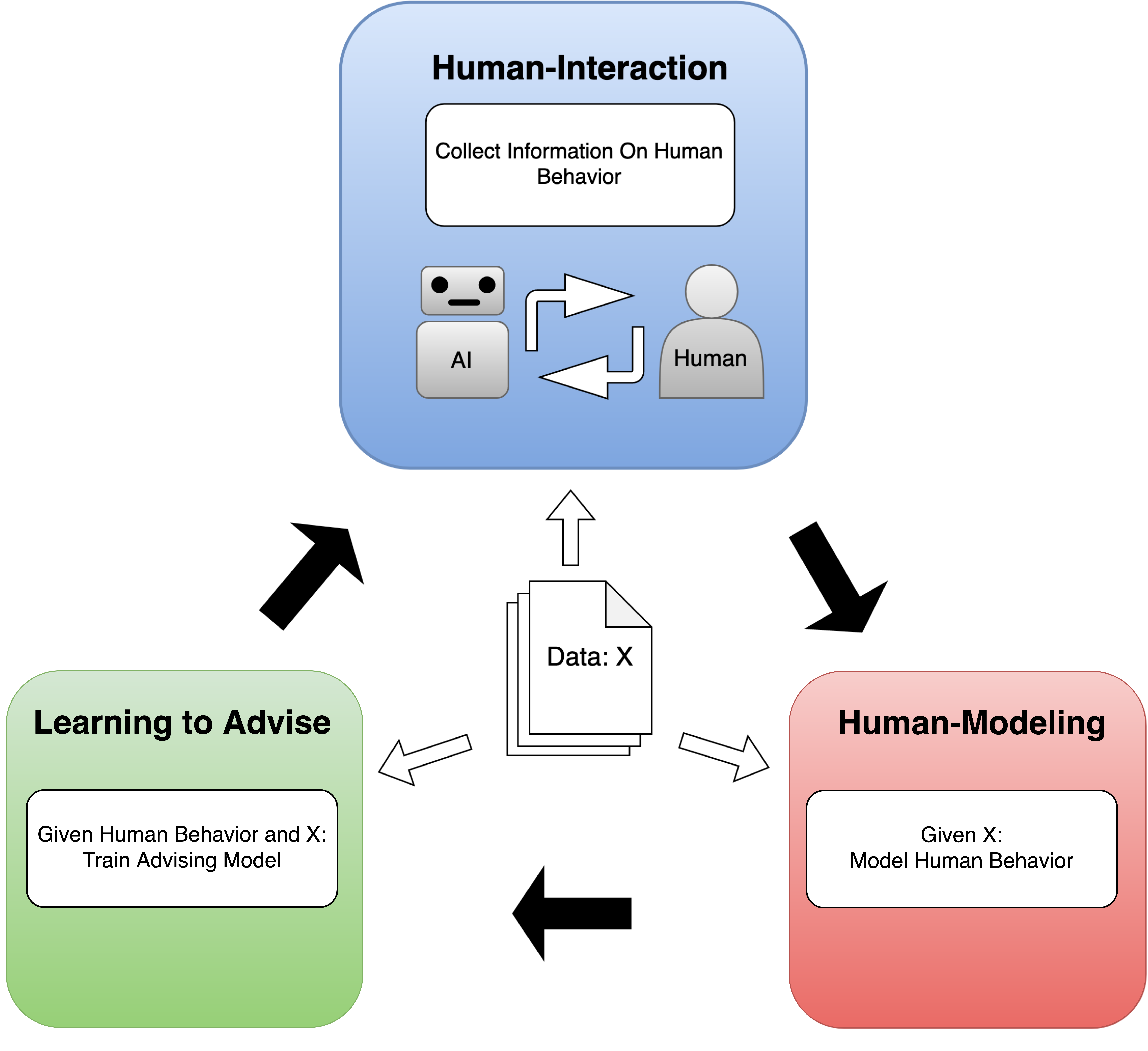}
    \caption{AI-assisted Team-Learning Framework}
    \label{fig:aidet}
\end{figure}

We propose the AI-assisted Team (AIaT)-Learning Framework, shown in Figure \ref{fig:aidet}. This framework includes three iterative phases that can be implemented in practice to develop HAI-team models that leverage a DMs behavior to benefit the team's performance. 
We first provide theoretical motivation for the potential benefits of leveraging ADB along with decision history and follow with a discussion on the phases of the AIaT-Learning framework. 


\subsection{Theoretical Motivation}\label{sec:motivation} 
We provide theoretical motivation for the potential benefits of leveraging ADB to improve team performance defined by an arbitrary loss function. Let $\mathcal{X}$ be the set of possible examples with labels $\mathcal{Y} = \{0,1\}$, where $\mathcal{X} \times \mathcal{Y} \sim {\mathcal{D}}$.
We define a human's ADB as their tendency to accept or reject algorithmic recommendations for arbitrary decision instances and feature values thereof. The human's ADB can be expressed as the probability $p(a|x)$ that the human would accept a contradicting recommendation for any given decision task defined by a feature vector   $x$. The event of accepting a contradicting recommendation is defined by the binary indicator variable $a$.

The overarching goal of our machine learning task is to learn a classifier $c^*: \mathcal{X} \rightarrow \mathcal{Y}$, where $c^* \in \mathcal{C}$ is selected by minimizing the expected loss $\mathcal{L}(y, c(x))$ as follows:
\begin{equation}
c^* = \underset{c \in \mathcal{C}}{\min} \big[\mathbb{E}_{x,y \sim \mathcal{D}} (\mathcal{L}(y, c(x)))\big ].
\end{equation}

However, for every instance defined by $x$, the classifier's recommendation $c(x)$ may get rejected by the human decision maker with probability $1-p(a|x)$. When a human rejects the classifier's recommendation, the classifier's recommendation $c(x)$ is replaced by the human's initial decision on the task $h$. We can thus obtain a classifier $c' \in \mathcal{C}$ by minimizing the loss from the final team decision: 

\begin{align}
  c' &= \nonumber \underset{c \in \mathcal{C}}{\min} \big[\mathbb{E}_{x,y \sim \mathcal{D}} (p(a|x)\mathcal{L}(y, c(x)) \\ & \indent + \nonumber (1-p(a|x))\mathcal{L}(y, h))\big] \\
  &= \nonumber \underset{c \in \mathcal{C}}{\min} \big[\mathbb{E}_{x,y \sim \mathcal{D}} (p(a|x)\mathcal{L}(y, c(x))\big] \\ & \indent + \nonumber \mathbb{E}_{x,y \sim \mathcal{D}}((1-p(a|x))\mathcal{L}(y, h)) \\ 
  &= \underset{c \in \mathcal{C}}{\min} \big[\mathbb{E}_{x,y \sim \mathcal{D}} (p(a|x)\mathcal{L}(y, c(x))\big]
\end{align}

In the above equation, we drop $\mathbb{E}_{x,y \sim \mathcal{D}}((1-p(a|x))\mathcal{L}(y, h))$ because it does not vary with choice of classifier $c \in \mathcal{C}$. By definition:

\begin{equation}
    \mathbb{E}_{x,y \sim \mathcal{D}} (p(a|x)\mathcal{L}(y, c'(x)) \leq \mathbb{E}_{x,y \sim \mathcal{D}} (p(a|x)\mathcal{L}(y, c^*(x))
\end{equation}

The above inequality demonstrates that $c'$ is a superior classifier in expectation because it directly minimizes expected loss over the instances that the human decision-maker would accept.

In practice, however, the human partner's ADB is unknown, and must be learned. We discuss how to obtain an estimate $\hat{p}(a|x)$ in the next section. 

\subsection{AI-assisted Team (AIaT)-Learning Framework} \label{sec:hai_framework}

Our AIaT-Learning framework consists of three phases. The \textit{Human Interaction Phase} serves as the data acquisition step during which we obtain information on the human partner's decisions and ADB. Given training data $\{X, Y\}$, we conduct two tasks involving the human. First, either historical data of the human's past decisions is obtained, or, in the absence of such history, the human records their decisions for a set of training instances; We refer to the resulting vector of the human's decisions as $H$. The second task involves acquiring data and modeling the human's ADB. Prior work established how a human's ADB can be predicted \citep{will_you_accept,CHONG2022107018}. In particular, prior work has found that a human's inherent self-reported confidence in their own decision, prior to receiving an algorithmic recommendation, is predictive of their ADB \citep{will_you_accept,CHONG2022107018}. In general, the greater the human's confidence in their own initial decision, the less likely they are to accept a contradictory algorithmic recommendation, independently of their confidence in the AI or the AI's explanation\footnote{One may consider that the human's ADB may be predicted exclusively from their decision history, given this history can predict their decision accuracy for a given instance. However, DMs' confidence, i.e., their assessment of their own accuracy, while shown to be predictive of their ADB, is rarely well-calibrated with respect to their true accuracy \citep{KLAYMAN1999216,green_disparate}.}. Thus, the human-interaction phase includes the acquisition of the DM's confidence in each of their decisions, denoted by vector $C$, for all training instances $X$. Additionally, following prior work on learning human's ADB \citep{will_you_accept,CHONG2022107018}, the human's decisions to accept or reject recommendations, denoted by $A$, are recorded whenever the human is presented with AI advice that contradicts their intitial judgement. 

In the \textit{Human-Modeling Phase}, the data acquired in the preceding steps is used to learn the \textit{discretion model} of the human partner's ADB. Specifically, as discussed above, prior work has established that humans' discretion outcomes are predictable given the human's confidence in their own decisions \citep{will_you_accept,CHONG2022107018}. 
Given that prior work established how to learn a mapping onto the human's discretion behavior, forming $\hat{p}(a|c,x)$, in this work we propose how this discretion behavior can be brought to bear towards learning to advise humans, so as to study its potential benefits. We leave the focus of developing superior discretion data acquisition and discretion model methods to future work, and we discuss related challenges in the Future Work section. For brevity, henceforth we denote $p(a|c,x)$ and $\hat{p}(a|c,x)$ as $p(a)$ and $\hat{p}(a)$, respectively.   
Additionally, our approach also brings to bear the human's decision behavior with respect to the underlying decision task, so as to complement the human's decision-making. In principle, this behavior can be directly observed in the historical data, as well as during deployment. In contexts where the human's decisions cannot be observed for all training instances in $X$, a model $\hat{h}(x)$ can be learned to infer the human's decisions (e.g., \citep{bansal_does_2021},\citep{madras_predict_2018}).


Finally, given training data $D = \{X, Y, H, \hat{p}(a)\}$, the \textit{Learning to Advise Phase} corresponds to simultaneously learning \emph{when} to advise the human and \emph{what} interpretable advice to offer by leveraging the human's decision history, discretion model, and tolerance for reconciliation costs, with the goal of optimizing the overall HAI team's performance. 
Our goal also entails defining the \emph{team performance} objective and metric that can reflect any given decision-making context. Next, we develop an algorithm for Learning to Advise.

\section{Learning-to-Advise Algorithm}
Here we develop TeamRules (\textsc{tr}), a rule-based algorithm which offers advice to a human decision-maker in the form of inherently interpretable rules. \textsc{tr} simultaneously learns \emph{when} to offer advice and \emph{what} advice to offer by leveraging the human's decision history, discretion model, and tolerance for reconciliation costs, so as to optimize the team's performance.
We begin with a formal definition of key terminology: 
\begin{definition}[Rule]
A \emph{rule} $r$ is a logical expression made of conditional statements about a subset of feature values. $r$ \emph{covers} an example $x_i$, denoted as $\mathbbm{C}(r, x_i) = 1$, if the logical expression evaluates to true given $x_i$. 
\end{definition}

\begin{definition}[Rule Set]
A \emph{rule set} $R$ is a collection of rules. $R$ \emph{covers} an example $x_i$, denoted $\mathbbm{C}(R, x_i) = \mathbbm{1}\{\sum_{r \in R} \mathbbm{C}(r,x_i) \geq 1\}$, if at least one rule $r$ in $R$ covers $x_i$.
\end{definition} 

\begin{figure}[t]
    \centering
    \includegraphics[width=0.775\columnwidth]{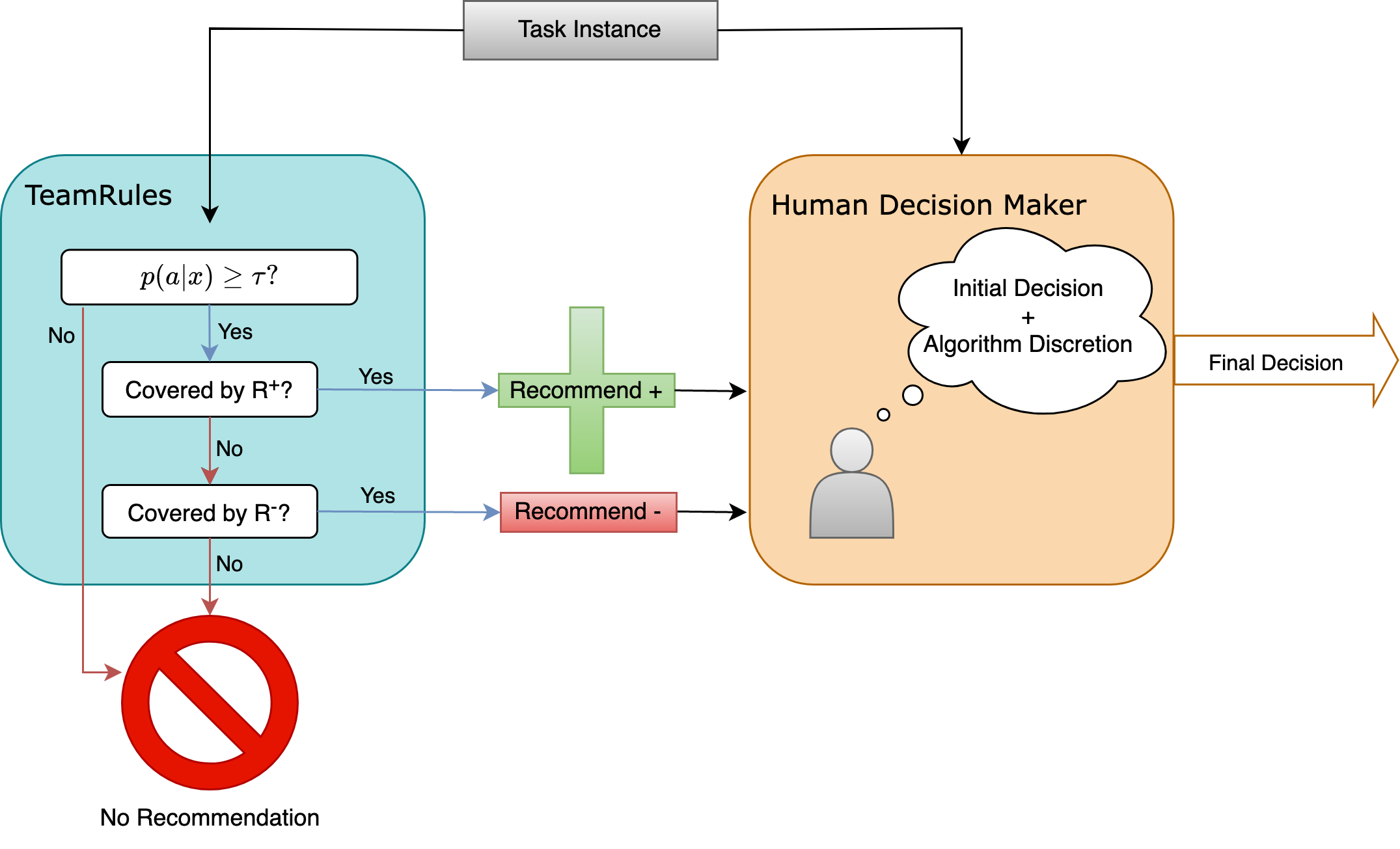}
    \caption{TeamRules + Human Partner Decision Process in AI-advised Decision-Making (AIaDM) Setting}
    \label{TR_deploy}
\end{figure}

Figure \ref{TR_deploy} illustrates the team's decision process in deployment when \textsc{tr} is paired with a human to produce decisions. \textsc{tr} takes as input the set of augmented instances $D = \{x_i, y_i, h_i, p(a_i)\}_{i=1}^n$, and then generates a set of positive and negative rules denoted $R^+$ and $R^-$, respectively, which predict $\hat{y_i}$ for instance $i$. For any given decision instance, \textsc{tr} provides a decision recommendation to the human if the instance is \emph{covered} by its rule set and if the advice for the instance is likely to be accepted according to the discretion model $\hat{p}(a)$, i.e., if $\hat{p}(a|x) \geq \tau = 0.5$; otherwise, \textsc{tr} does not advise the human on the corresponding decision instance. 

The \textsc{tr} decision-making process in training is specified as follows:
\begin{align} \label{dec_rule}
   \nonumber &\textbf{if}: \mathbbm{C}(R^+, x_i), \text{ then } \hat{y_i} = 1\\
   &\textbf{elif}: \mathbbm{C}(R^-, x_i), \text{ then } \hat{y_i} = 0\\
   \nonumber &\textbf{else}: \text{provide no recommendation (use $h_i$)}  
\end{align}
 
The Team Loss objective is given by $\mathcal{L}(D, R)$:
\begin{equation} \label{eq:utility2}
\begin{aligned}
    \mathcal{L}(D,R) &=  \ell(D,R) + \omega(D,R) \\
    &= \sum_{i=1}^n\big[p(a_i)\mathbbm{1}(y_i \neq \hat{y_i}) + \alpha\mathbbm{1}(\hat{y_i} \neq h_i)\big]
\end{aligned}
\end{equation}

Where $\hat{y}_i$ is the HAI team's decision, produced by the decision process given in Eq. \ref{dec_rule},  $\ell(D, R) = \sum_{i=1}^n\big[p(a_i)\mathbbm{1}(y_i \neq \hat{y_i})\big]$ is the expected (discretion-adjusted) team decision loss, $\omega(D, R) = \alpha\sum_{i=1}^n\mathbbm{1}(\hat{y_i} \neq h_i)$ is the total reconciliation cost, where $\alpha$ reflects the human partner's reconciliation cost incurred from reconciling a single contradictory recommendation. Note that $\alpha$ reflects the human's cost of reconciling a contradictory recommendation relative to the cost of an incorrect decision. In Appendix \ref{app:objective} we include a detailed formulation of the objective which does not reference Eq. \ref{dec_rule}. 

Intuitively, $\ell(D, R)$ allows our method to prioritize recommendations that are more likely to be accepted by the human because the training instances are weighted by the likelihood with which they will get accepted. By penalizing contradictions, $\omega(D, R)$ allows our method to reflect the context-specific (e.g., the human DM's) tolerance to incurring additional reconciliation costs for superior team decision accuracy. An $\alpha = 0$ indicates that recommendations which contradict the human's judgment can be provided to the human if there is any chance they will decrease team decision loss with no regard for the reconciliation burden placed on the human partner. Larger values of $\alpha$ indicate the human is less tolerant of contradictions to benefit the team's decisions; thus, recommendations ought to be provided more \emph{selectively}. Finally, at $\alpha = 1$, our model should provide no recommendations because any improvement in decisions enabled by a recommendation (correcting a mistake) is offset by the cost incurred by the human's reconciliation of the AI's recommendation. 

The task of the algorithm is thus to generate from the input data a set of optimal rules $R^*$, where $R = R^+ \bigcup R^-$ and $
    R^* \in \underset{R}{\text{argmin }} \mathcal{L}(D, R)$.

\subsection{TeamRules Optimization}
\begin{algorithm}[t]
\caption{TeamRules}\label{alg:team}
\begin{algorithmic}[1]
\State \textbf{Input:} $D$
\State \textbf{Parameters: $T, \alpha, C_0, \beta_0, \beta_1, \beta_2, q$} 
\State $\Gamma^+ \gets \text{RF-Generated Positive Candidate Rule Set}$
\State $\Gamma^- \gets \text{RF-Generated Negative Candidate Rule Set}$
\State $R_0 \gets \emptyset$
\State $R^* \gets R_0$
\For{$t=1...T$}
\State $R_t \gets R_{t-1}$
\For{$i=0...n$}
\State $\phi_i \gets \mathcal{L}(x_i, R_t)$ \Comment{$\mathcal{L}$ defined in Eq. \ref{eq:utility2}}
\EndFor
\State $\epsilon \gets sample(i | \phi)$ \Comment{Sample an instance. The likelihood of selecting instance $i$ is proportional to $\phi_i$.}
\If{$\mathbbm{C}(R_{t-1},x_\epsilon)$ and ($\hat{y}_\epsilon \neq y_\epsilon$) or ($h_\epsilon \neq \hat{y}_\epsilon$)}
\If{$(y_{\epsilon} = 0)$}
\State $R_{t} \gets$ cut rule from $R^+_t$
that covers $x_\epsilon$
\Else{}
\If{$randint(0,1)$}
\State $R_{t} \gets$ add rule to $R_{t}^+$ to cover $x_\epsilon$ 
\Else{}
\State $R_{t} \gets$ cut rule from $R_{t}^-$ that covers $x_\epsilon$
\EndIf
\EndIf
\Else{}
\State $R_{t} \gets$ add rule to $R^{sign(y_\epsilon)}_{t-1}$
\EndIf
\If{$\mathcal{L}(D, R_{t}) < \mathcal{L}(D, R_{t-1})$}
\State $R^* \gets R_{t}$
\EndIf
\If{$\exp \left( {\frac{\mathcal{L}(D, R_{t-1}) - \mathcal{L}(D, R_{t})}{C_0^{\frac{t}{T}}}} \right) \leq random()$}
\State $R_t \gets R_{t-1}$ 
\EndIf
\EndFor
\State \textbf{Output:} $R^*$
\end{algorithmic}
\end{algorithm}

We develop an algorithm to construct a \textsc{tr} model that  optimizes the team loss objective using a simulated annealing procedure. Simulated annealing is suitable given our discrete space, and has been used successfully in previous rule-based models  \citep{sim_anneal, wang2017bayesian}. Our algorithm takes as input data $D$ and begins by generating a set of $\beta_0$ candidate rules $\Gamma$, with maximum rule length $\beta_1$ and minimum support of each rule $\beta_2$. $\Gamma$ represents the complete set of rules from which rules can be selected to optimize the team's objective. $\Gamma$ is generated using an off-the-shelf rule mining algorithm, such as FP-growth \citep{wang_gaining_nodate}. Next, the starting solution, $R_0$, is set to an empty set of rules, i.e., decisions are undertaken exclusively by the human. 

At each iteration $t$ the current rule set $R_t$ is adapted such that the team loss $\mathcal{L}(D,R_t)$, shown in Eq \ref{eq:utility2}, is progressively minimized. First, \textsc{tr} randomly draws a single instance $\epsilon$, such that the probability that an instance is selected is proportional to its contribution to the team's loss, given by $\mathcal{L}(x_i, R_t)$. Subsequently, the current rule set $R_t$ is adapted to correct the team's decision for the sampled instance, either by adding or removing rules. Specifically, if $\epsilon$ is currently covered by $R_t$ and is incorrectly classified (increasing decision loss) or incurs a reconciliation cost, \textsc{tr} attempts to reduce the team loss either by removing a rule that is covering it incorrectly or by adding a rule that would lead to correct decision. If $\epsilon$ is not covered by $R_{t}$, an appropriate rule is added to cover the instance. Once $R_t$ has been adapted, if $\mathcal{L}(D, R_{t}) < \mathcal{L}(D, R_{t-1})$, we set $R^* \leftarrow R_{t}$. Finally, to encourage exploration, we reset $R_t \leftarrow R_{t-1}$ with probability:
$1-\exp \left( {\frac{\mathcal{L}(D, R_{t-1}) - \mathcal{L}(D, R_{t})}{C_0^{\frac{t}{T}}}} \right)$. Note that the probability of resetting to the previous rule set decreases over time. The complete algorithm is shown in Algorithm \ref{alg:team}.

The process above leads to a set of rules uniquely adapted to complement the human partner. The process of updating the rule sets to minimize the team loss objective can be viewed as selecting which instances would be covered and on which to advise the human; in addition, the rules are customized to focus on instances which the human is likely to both err on and accept recommendations for. As $\alpha$ increases, correcting the human via contradiction is less rewarding to the team's objective, given contradicting the human is more costly; the model thus provides fewer recommendations, limited to instances that the human is most likely to accept recommendations for and err on. Finally, given the human can have different likelihoods of accepting a AI advice for different instances covered by a given rule, a recommendation for an instance covered by the rules is provided if the advice is likely to be accepted according to the discretion model $\hat{p}(a)$, i.e., if $\hat{p}(a|x) > 0.5$. 


\section{Experiments}
We evaluate \textsc{tr}'s performance relative to existing, inherently interpretable rule-based methods across different AIaDM settings. We present results for simulated and real-world datasets, and consider different human behaviors and reconciliation costs to understand how these aspects of a setting affect \textsc{tr}'s team performance.

\paragraph{Benchmarks}
To our knowledge, no existing method leverages a human partner's ADB and decision history to learn how to selectively advise a high-stakes DM with rule-based recommendations that maximize team decision accuracy and minimize reconciliation costs.
Consequently, we utilize the inherently interpretable rule-based Bayesian Rule Set (\textsc{brs}) method, which does not leverage any human information \citep{wang2017bayesian}, to assess whether \textsc{tr} yields better team performance than can be achieved by advising the human with rules optimized solely for predictive accuracy with no consideration of the human. In addition, we examine whether leveraging the human's discretion model and regularizing for reconciliation costs is advantageous for \textsc{tr}'s team performance by adapting the existing \textsc{\normalsize{h}\scriptsize{y}\normalsize{rs}} rule-based method as a benchmark. While \textsc{\normalsize{h}\scriptsize{y}\normalsize{rs}} was initially designed to provide partial interpretability for a black-box model, we propose a novel use for the method in which the black-box model is replaced with a human decision, allowing \textsc{\normalsize{h}\scriptsize{y}\normalsize{rs}} to function as an advising model that leverages the human's decision history. Unlike \textsc{tr}, however, \textsc{\normalsize{h}\scriptsize{y}\normalsize{rs}} does not take the human's ADB or tolerance for reconciliation costs into account. 

We compare all three methods in the AIaDM setting: the rule-based model makes recommendations, and a human partner decides whether to accept them. To assess the value of leveraging a discretion model independently of the discretion model's accuracy, we first assume perfect knowledge of the human's ADB. We later assess \textsc{tr}'s performance as discretion model accuracy decreases. 
 
\paragraph{Datasets}
We experiment on three real-world and two synthetic datasets. The real-world decision-making tasks we utilize include the FICO dataset \citep{fico_2018} for making loan decisions, an HR employee attrition dataset that informs retention decisions \footnote{https://www.kaggle.com/datasets/pavansubhasht/ibm-hr-analytics-attrition-dataset}, and the Adult dataset from the UCI Machine Learning Repository \citep{uci_adult}. The FICO and HR dataset represent high-stakes settings in which the outcome of the decision task has a significant impact on individuals' lives. Following the convention of many rule-based models \citep{wang2017bayesian,angelino2018learning}, we limit the model capacity for these datasets to a rule length of 3 to ensure all models remain easy to interpret by a DM and are compared under the same capacity limits. The first simulated dataset (named Checkers) is a 2-dimensional checkerboard example in which we limit the model capacity to a rule length of $1$. This limit on model capacity is set to create an exaggerated scenario in which it is impossible for any model in this class to get higher than 50\% accuracy without leveraging additional human information. The second dataset (named Gaussian) is a more complex, 20-dimensional dataset in which the true underlying classification mapping is a complex function of all features. The synthetic datasets allow for a clear analysis of results in a controlled environment. 

\paragraph{Simulating Human Algorithm Discretion Behavior}
We evaluate our method's performance under a variety of human behaviors. We first simulate human \emph{decision} history such that the human has varying decision accuracy across instances by partitioning the data into high and low human accuracy regions using an arbitrary function of all available features.  In addition, we evaluate \textsc{tr} for three different simulated human ADBs: \emph{Rational}, \emph{Neutral}, and \emph{Irrational}. A \textit{rational} DM knows their relative decision accuracy for different instances and accepts recommendations in the region in which they have low accuracy while rejecting in the region they have high accuracy. An \textit{irrational} DM does the opposite: they irrationally reject recommendations when they have low accuracy and accept recommendations when they have high accuracy. Lastly, we simulate a \textit{neutral} human DM who is often, but not always, rational. The neutral behavior may better reflect human experts, as humans rarely exhibit either perfectly rational or irrational discretion \citep{JOHNSON2021203}. We simulate the neutral human behavior such that most, but not all, of the instances the human accepts recommendations for are instances in the human's low accuracy region. The neutral human also accepts recommendations for some instances in their high accuracy region. Given the above behavior, the human exhibits deterministic ADB to accept or reject recommendations. A stochastic behavior is likely in practice and primarily affects the predictive accuracy of the discretion model that predicts the human's ADB, and thereby \textsc{tr}'s performance. 
We thus also evaluate our approach with discretion models of varying accuracy. 

We vary discretion model accuracy by training a discretion model using an out-of-the-box classifier (XGBoost) \citep{xgboost} while varying the number of training observations used. Model accuracy decreases as the number of training observations used decreases. An observation $i$ is described by features $x_i$ and label $a_i$, where $a_i$ is the human's simulated response to receiving an algorithmic recommendation. Note that this process serves as a simulation of the Human-Modeling phase in which it could be difficult to obtain a large number of DM responses to algorithmic recommendations. We provide full details on our datasets and human behavior simulation procedures in Appendix \ref{exp_deets}. 

\paragraph{Evaluation Metrics}
Our primary goal is to evaluate the Total Team Loss (TTL) on the test set $\tilde{D}$. TTL includes the team's decision loss and the loss from incurred reconciliation costs. The team's decision loss (TDL) on test data $\tilde{D}$ is given by: $
\textstyle
    \text{TDL} = ({\sum}_{i \in \tilde{D}}|\hat{y}_i - y_i|) / N_{\tilde{D}}$.
\normalsize
The total contradiction loss (CL) is given by: $
\textstyle
    \text{CL} = (\alpha\sum_{i \in \tilde{D}}|\hat{y}_i - h_i|)/{N_{\tilde{D}}},
$
\normalsize
where $\alpha$ is the cost of reconciling a single contradicting recommendation and $N_{\tilde{D}}$ is the total number of observations in test set $\tilde{D}$. Finally, we define the total team loss (TTL) as: $
\small \textstyle
    \text{TTL} =\text{TDL} +\text{CL}
$\normalsize.

\section{Results}

We state key questions followed by discussion on the experiments and results which address them.

\begin{figure}[t]
    \centering

        \begin{subfigure}[b]{0.745\columnwidth}  
            \centering 
            \includegraphics[width=\textwidth]{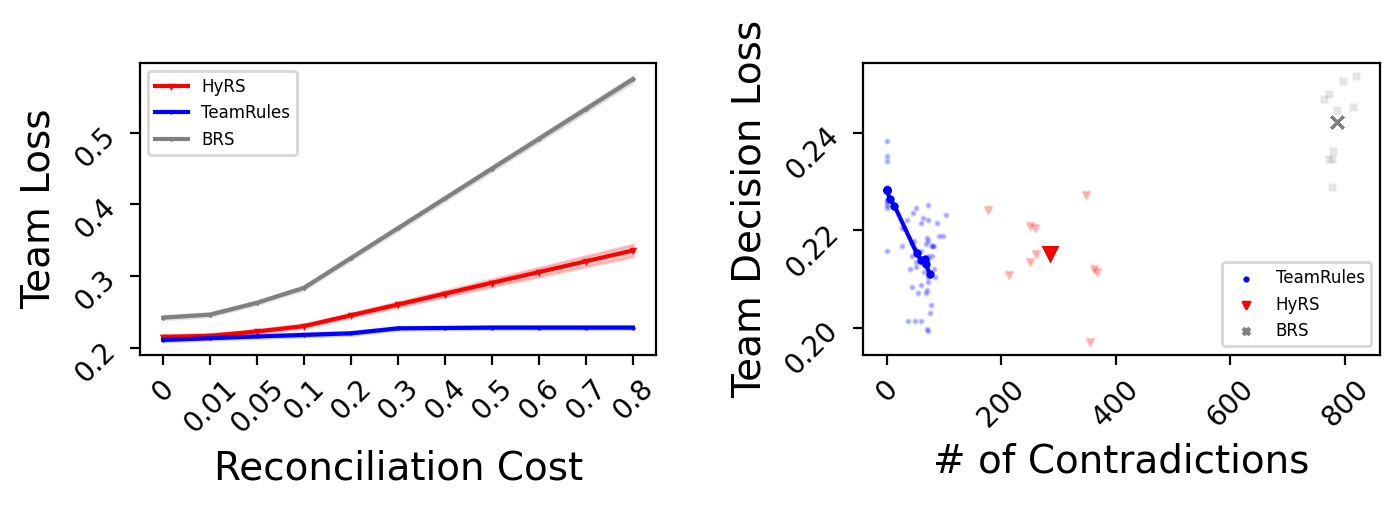}
            \caption[]%
            {{\small Neutral}}    
            \label{fig:mean and std of net24}
        \end{subfigure}
        \hfill
        \begin{subfigure}[b]{0.745\columnwidth}
            \centering
            \includegraphics[width=\textwidth]{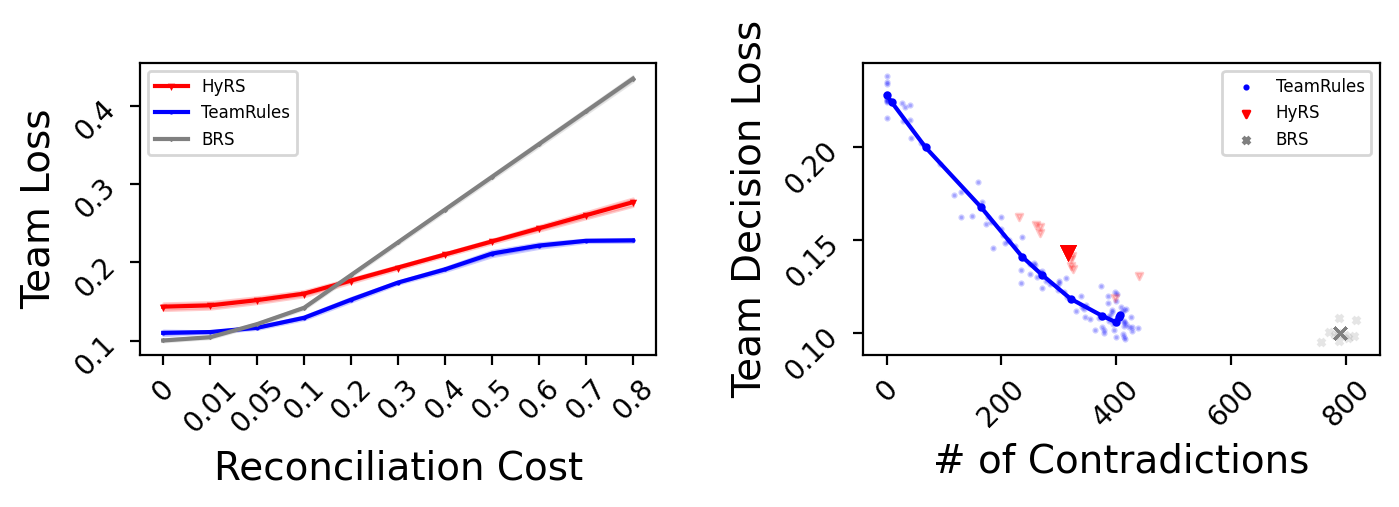}
            \caption[]%
            {{\small Rational}}    
            \label{fig:mean and std of net14}
        \end{subfigure}
        \hfill
        \begin{subfigure}[b]{0.745\columnwidth}   
            \centering 
            \includegraphics[width=\textwidth]{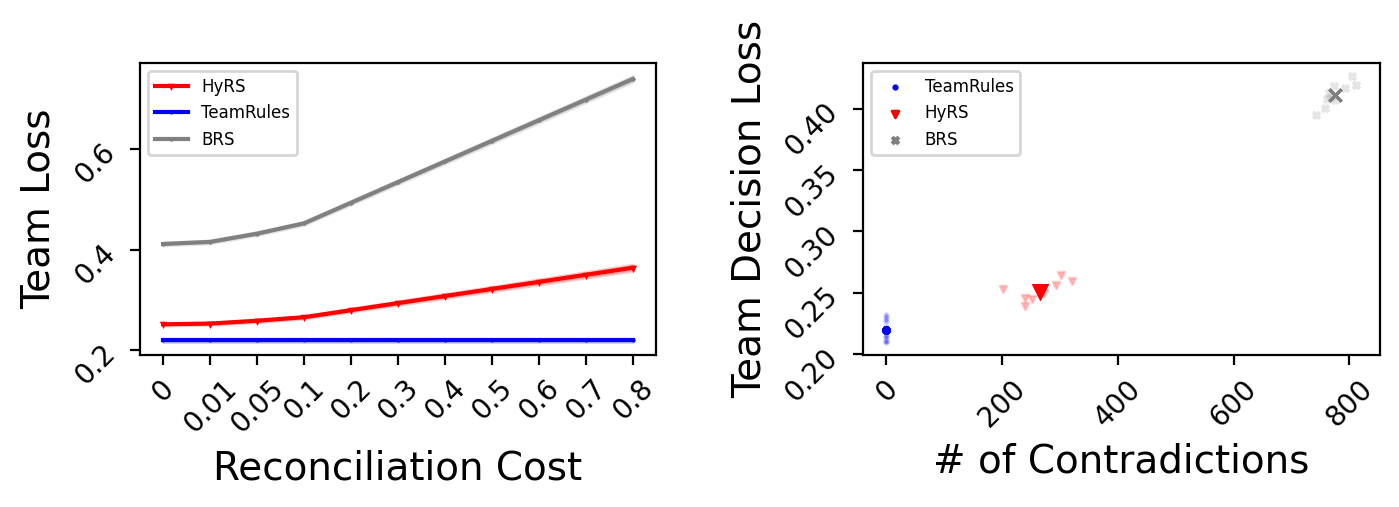}
            \caption[]%
            {{\small Irrational}}    
            \label{fig:mean and std of net44}
        \end{subfigure}
\caption[]{Total team loss for varying reconciliation costs. (Left) Total Team Loss vs. reconciliation cost. Results show average $+-$ SE. (Right) TDL vs. Number of contradictions. Line connects averages for different cost settings. Transparent points show individual results for cost$+$run.}
\label{fig:main} 
\end{figure}

\subsection{Does \textsc{tr} lead to lower team loss relative to benchmarks as reconciliation costs vary?} 
We first compare the performance of \textsc{tr} relative to the benchmarks across a variety of reconciliation costs. Different environments are characterized by the human's unique tolerance for incurring additional reconciliation costs to improve the team's decisions, and we thus evaluate \textsc{tr} and \textsc{\normalsize{h}\scriptsize{y}\normalsize{rs}} for different degrees of tolerance by varying the reconciliation costs from 0 to 0.8. Here we also aim to assess \textsc{tr}'s benefits from bringing to bear the human's ADB, towards which we assume that the human's ADB can be correctly estimated; we later study how any benefits vary as the estimation of the human's ADB degrades. 

Results for the FICO dataset are shown in Figure \ref{fig:main} and all other results are consistent and shown in Appendix \ref{all_results}. The plots on the left of Figure \ref{fig:main} show total team loss (TTL) as a function of the reconciliation cost incurred by the DM from a single contradicting recommendation\footnote{Recall that lower reconciliation costs indicate that correcting a human's decision through advising is valued more than avoiding a contradiction; when the cost is 1, the value from correcting a decision is entirely offset by the additional reconciliation cost incurred}. As shown in Figure \ref{fig:main}, \textsc{tr} leads to lower total team loss in most settings and is otherwise comparable to the best alternative. 

Figure \ref{fig:main} shows that when reconciliation costs rise, the gap in Total Team Loss between \textsc{tr} and the other two baselines increases. This is because
as reconciliation costs rise, there are more opportunities to benefit the team by leveraging the DM's ADB and via accounting for the human DM's reconciliation costs.  In addition, \textsc{brs} leads to significantly higher team loss, followed by \textsc{\normalsize{h}\scriptsize{y}\normalsize{rs}}. Recall that \textsc{brs} always make recommendations: it does not not bring to bear the human's expertise and ADB, and thus it overburdens the DM when reconciliation costs are high. \textsc{\normalsize{h}\scriptsize{y}\normalsize{rs}} is able to make recommendations selectively based on its ability to leverage the human's decision history and can outperform \textsc{brs} when reconciliation costs are high because it does not always make a recommendation to the human. Although \textsc{\normalsize{h}\scriptsize{y}\normalsize{rs}} is able to withhold recommendations leading to fewer contradictions, it significantly lags behind \textsc{tr} in terms of team loss. 

By leveraging the human DM's decision history, ADB, and tolerance for reconciliation costs, \textsc{tr} is able to either improve on the human's standalone performance or withhold offering advice to the human when the advice is likely to harm the DM's standalone performance. Our results show that \textsc{tr} effectively leverages the human's ADB, given its differential performance that can be explained by \textsc{tr}'s accounting for the accept and reject regions.  Specifically,  \textsc{tr} would never generate rules which might improve performance in the reject region to the detriment of performance in the accept region. It is also able to discern that contradictions in the reject region harm team performance, given they will give rise to reconciliation costs without improving the final team decision. \textsc{tr}'s goal is to find the optimal set of rules in the human's accept region; if it is unable to find a set of rules that improves on the human's standalone performance in the accept region, it does not offer advice to the human. \textsc{\normalsize{h}\scriptsize{y}\normalsize{rs}} and \textsc{brs} are unable to accomplish this because they might find rules which appear to improve team loss when evaluated on all instances, however, the DM does not accept any recommendations in the reject region and the rules do not lead to better outcomes in the accept region. 

When reconciliation costs are low or 0,  withholding recommendations does not benefit  the team's performance and the only way for an AI to benefit the team is to identify a set of rules which lead to superior decision accuracy on the \emph{accept} region. As shown in Figure \ref{fig:main}, \textsc{tr} is indeed able to do so and improves team performance relative to \textsc{brs} and \textsc{\normalsize{h}\scriptsize{y}\normalsize{rs}} across data sets by specializing in producing superior rules for the accept region.


The complete set of results for when there are no reconciliation costs is shown in Table \ref{tabr}. In all irrational settings with reconciliation costs set to 0, the human is perfect on the accept region, so the best course of action is to allow the human to undertake all the decisions. This is because recommendations in the reject region will not be accepted, and recommendations in the accept region cannot be better than the human's decision. As shown, \textsc{tr} effectively  identifies that it should not advise the DM in this context. \textsc{brs} does not withhold recommendations and thereby harms the DM's performance, while \textsc{\normalsize{h}\scriptsize{y}\normalsize{rs}} is occasionally able to identify that it cannot improve on the human in the accept region. If \textsc{\normalsize{h}\scriptsize{y}\normalsize{rs}} identifies a set of rules that provide more benefit on the DM's reject region than cause harm on the accept region, it  will provide advice that will be rejected. 

As shown in Table \ref{tabr}, in the neutral settings with reconciliation costs set to 0, \textsc{tr} remains the method of choice. In this setting, the human exhibits imperfect discretion of the AI's advice and \textsc{tr}'s accounting for this behavior allows it to focus on producing superior advice for instances in the accept region, leading to its consistent advantage in this setting. Recall that \textsc{tr} was especially motivated by the challenge posed by imperfect human discretion of the AI's advice. Our results show that both when the human is neutral and entirely irrational, \textsc{tr} is consistently the method of choice.

In the rational setting with 0 reconciliation costs, it is optimal to offer the human advice for all instances, because the human exhibits optimal AI discretion and always accepts advice that improves the team's performance. In this context, we see an example for the cost of data-driven learning. Recall that \textsc{brs} offers advice for all instances and does not aim to learn from data how to selectively advise a human. While this strategy is harmful when the human discretion is imperfect and when the human incurs reconciliation costs, it is optimal when the human discretion is optimal and there are no reconciliation costs to the human. Consequently, \textsc{brs} often yields the lowest loss in this setting. \textsc{\normalsize{h}\scriptsize{y}\normalsize{rs}} and \textsc{tr}, on the other hand, aim to learn from the historical data on what instances it is beneficial to advise the human, resulting in higher loss in the Gaussian, Adult, and FICO data sets. 

Overall, our results demonstrate that when the human exhibits imperfect discretion of the AI's advice, \textsc{tr} can advantageously bring to bear the human's discretion behavior to both produce superior advice for instances the human is likely to accept advice for and to selectively offer such advice. In addition, we find that \textsc{tr} effectively trades-off reconciliation costs with benefits to decision-making that reflect the environment's (human's) tolerance for reconciliation costs to produce lower overall team loss.



\begin{table}[t] 
\centering
\caption{Average Total Team Loss (TTL) over all repetitions when reconciliation cost is 0. \emph{Human} column shows human's standalone decision loss prior to receiving recommendations. Paired sample one-sided t-test used to compare \textsc{tr} to \textsc{\normalsize{h}\scriptsize{y}\normalsize{rs}} ($^*:p<0.05$, $^{**}:p<0.005$) and \textsc{tr} to \textsc{brs} ($_\diamond: p<0.05$, $_{\diamond\diamond}:p<0.005$).}
\normalsize
\begin{tabular}{l|cc|ccc}
\hline
Data & ADB & Human & \textsc{tr} & \textsc{\normalsize{h}\scriptsize{y}\normalsize{rs}} & \textsc{brs} \\ \hline
Checkers & Rational   & $ .093 $ & $ .063^{**}_{\diamond\diamond} $        & $ .093 $ & $ .128 $ \\
         & Neutral    & $ .093 $ & $ .084^{*}_{\diamond} $   \space          & $ .093 $ & $ .177 $ \\
         & Irrational & $ .103 $ & $ .103_{\diamond\diamond} $             & $ .103 $ & $ .230 $ \\ \hline
Gaussian & Rational   & $ .251 $ & $ .211^{*} $   \space                         & $ .224 $ & $ .181 $ \\
         & Neutral    & $ .251 $ & $ .250_{\diamond\diamond}$              & $ .251 $ & $ .264 $ \\
         & Irrational & $ .258 $ & $ .258^{**}_{\diamond\diamond} $        & $ .267 $ & $ .464 $ \\ \hline
Adult    & Rational   & $ .171 $ & $ .137^{*} $  \space                          & $ .143 $ & $ .128$  \\
         & Neutral    & $ .171 $ & $ .170_{\diamond\diamond} $             & $ .170 $ & $ .173 $ \\
         & Irrational & $ .172 $ & $ .172^{**}_{\diamond\diamond} $        & $ .184 $ & $ .234 $ \\ \hline
FICO     & Rational   & $ .229 $ & $ .111^{**} $                           & $ .144 $ & $ .100 $ \\
         & Neutral    & $ .229 $ & $ .211^{*}_{\diamond\diamond} $         & $ .215 $ & $ .241 $ \\
         & Irrational & $ .220 $ & $ .220^{**}_{\diamond\diamond} $        & $ .252 $ & $ .412 $ \\ \hline
HR       & Rational   & $ .373 $ & $ .101_{\diamond\diamond}$ & $ .083 $ & $ .144$  \\
         & Neutral    & $ .373 $ & $ .364 $\space\space\space                & $ .362 $ & $ .381 $ \\
         & Irrational & $ .351 $ & $ .351^{**}_{\diamond\diamond} $        & $ .409$  & $ .435 $ \\ \hline
\end{tabular}
\normalsize
\label{tabr}
\end{table}
\normalsize

\subsection{How does \textsc{tr} improve over \textsc{h}\textsc{\normalsize{y}}\textsc{rs}?}
We now focus on the plots on the right of Figures \ref{fig:main}a,b,c that show the team decision loss (TDL) vs. the number of contradictions. These plots correspond to those on the left of Figure 3 and are a different visualization of the same result. In the plots on the right, we see how the number of contradictions and decision accuracy of each method varies as the reconciliation cost varies. Unlike the plots on the left, the plots on the right demonstrate model performance without assuming that the reconciliation cost parameter represents a ground truth of the setting, rather, it is used as a means to control the number of contradictions \textsc{tr} is allowed to make without explicitly translating contradictions to a team loss. 

In the neutral and rational settings in Figure \ref{fig:main} (a) and (b), we observe that by varying the reconciliation cost parameter in the \textsc{tr} model, we can trade contradictions for TDL, generating a curve. While \textsc{tr} is not always able to identify the same set of rules as \textsc{brs} with a reconciliation cost parameter of 0, it is still able to achieve very similar performance in terms of team decision performance albeit with much fewer contradictions. Similarly, we observe that \textsc{tr} is able to achieve equivalent or lower TDL with the same number of contradictions as \textsc{h}\textsc{\normalsize{y}}\textsc{rs} and it is able to achieve the same TDL with the same or fewer contradictions compared to \textsc{h}\textsc{\normalsize{y}}\textsc{rs}. In the irrational settings, we can see that \textsc{tr} leads to no contradictions and has performance fixed to that of the human, while \textsc{h}\textsc{\normalsize{y}}\textsc{rs} and \textsc{brs} make contradictions which harm TDL. Even though \textsc{tr} does not always lead to lower TDL, it is always more efficient by leading to similar team decision loss with much fewer contradictions. 

\begin{figure}[t]
    \centering
        \begin{subfigure}[b]{0.32\columnwidth}  
            
            \includegraphics[width=\textwidth]{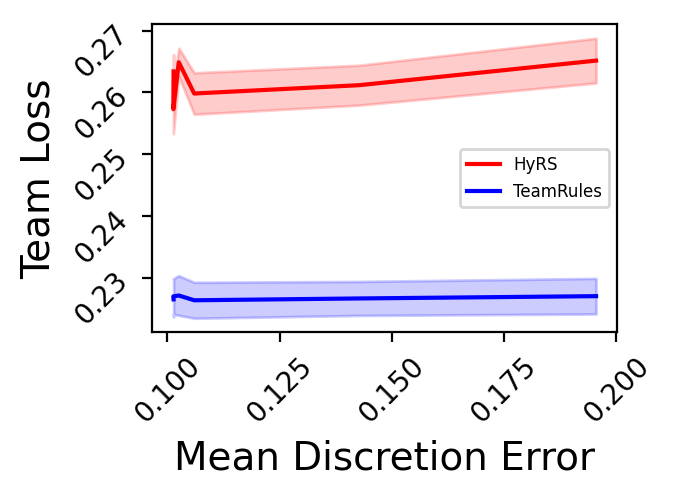}
            \caption[]%
            {{\small Neutral}}    
        \end{subfigure}
        \begin{subfigure}[b]{0.32\columnwidth}
           
            \includegraphics[width=\textwidth]{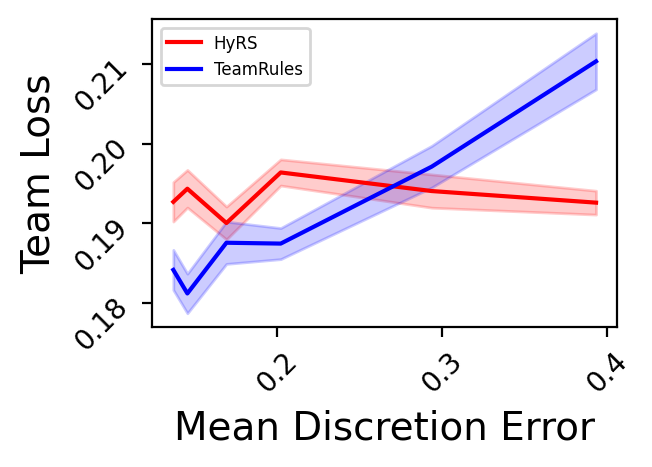}
            \caption[]%
            {{\small Rational}}    
        \end{subfigure}
        \begin{subfigure}[b]{0.32\columnwidth}   
            
            \includegraphics[width=\textwidth]{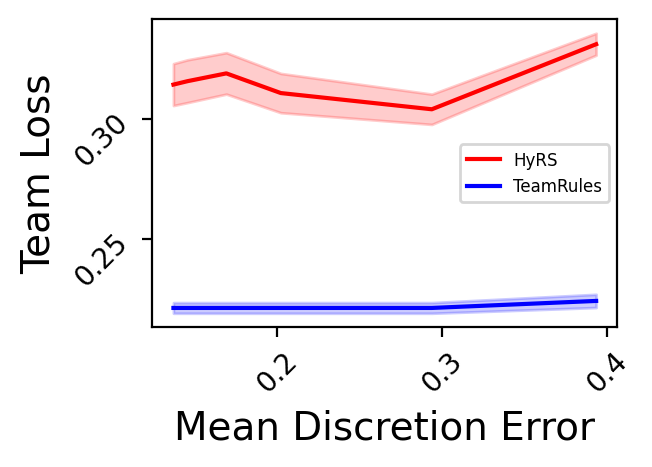}
            \caption[]%
            {{\small Irrational}}    
        \end{subfigure}
    \caption{Total Team Loss on FICO data with 0.3 reconciliation cost. Discretion model learned using XGBoost on subsets of data leading to decreasing discretion model accuracy. Avg. of 5 reps. with randomized 80-20\% train-test split on each run.}
    \label{fico_err}\vspace{-5mm}
\end{figure}

\subsection{How robust is \textsc{tr} to inaccurate discretion models?}
In real-world applications, we do not assume access to a perfectly accurate discretion model. In practice, we would estimate ADB by training a model on a small set of labeled data, obtaining $\hat{p}(a)$ with imperfect accuracy. We investigate how the \emph{accuracy} of $\hat{p}(a)$ impacts team performance and how much it can deteriorate before \textsc{tr} no longer provides value. We gradually decrease the accuracy by simulating the \emph{Human-Modeling Phase} of the AIaT-Learning Framework with limited observations of human behavior, as would be typical in practice. Using the simulated human behavior for each setting, we observe whether the DM accepts recommendations in the \emph{Human-Interaction} phase. Next, we take increasingly limited subsets of these observations and use the subsets to train discretion models using an out of the box classifier such as XGBoost. As the number of observations used in training decreases, the discretion model accuracy also decreases, allowing us to simulate the entire process with varying discretion model accuracies. 

Results for the FICO dataset, assuming a true reconciliation cost of $0.3$, are shown in Figure \ref{fico_err} (results for Gaussian data are shown in Appendix \ref{all_results}). We find that while \textsc{tr} performance may deteriorate with decreasing discretion model accuracy, it generally still outperforms \textsc{\normalsize{h}\scriptsize{y}\normalsize{rs}} with model accuracies of 60\%-85\%, depending on the setting.  Intuitively, small errors in $\hat{p}(a)$ cause \textsc{tr} to place a small weight on some instances in $\mathcal{D}_\mathcal{R}$. In practice, learning from a few out-of-distribution examples would not necessarily hurt the model. As the discretion model accuracy decreases, \textsc{tr} randomly focuses on instances from the entire data set. Finally, the more benefit \textsc{tr} provides under the assumption of a perfect discretion model, the more discretion model accuracy can decrease before \textsc{tr} no longer provides benefit.

\subsection{What is the value of the selective recommendation mechanism?}
\begin{figure}[t]
    \centering
    \begin{subfigure}[b]{0.725\columnwidth}  
            \centering 
            \includegraphics[width=\textwidth]{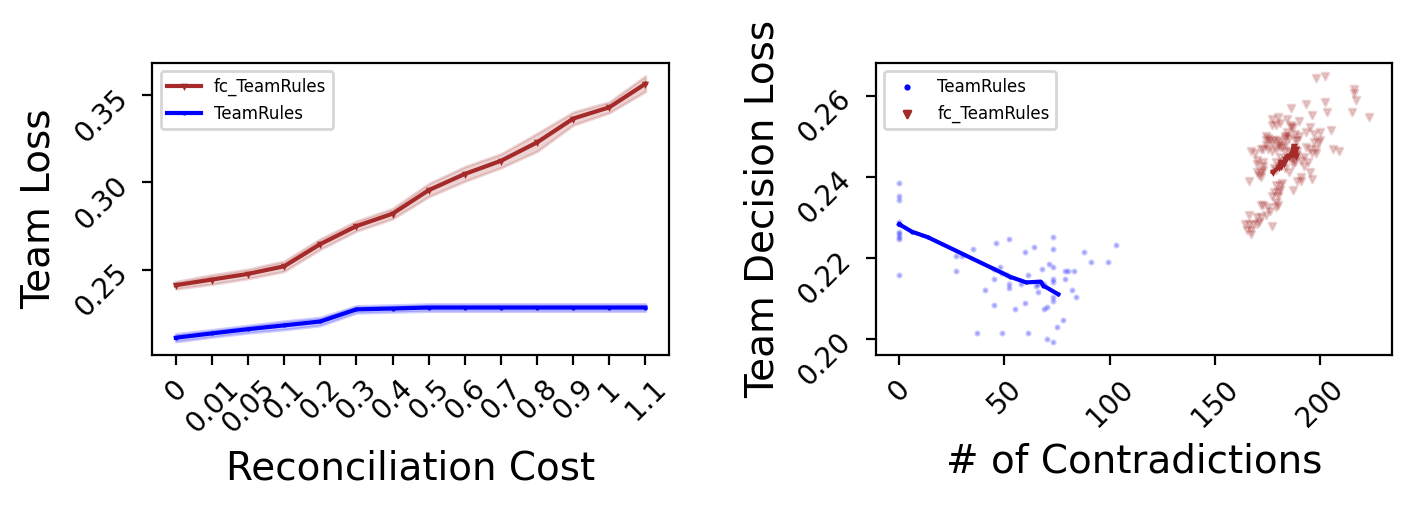}
            \caption[]%
            {{\small Neutral}}    
            \label{fig:mean and std of net24}
        \end{subfigure}
        \hfill
        \begin{subfigure}[b]{0.725\columnwidth}
            \centering
            \includegraphics[width=\textwidth]{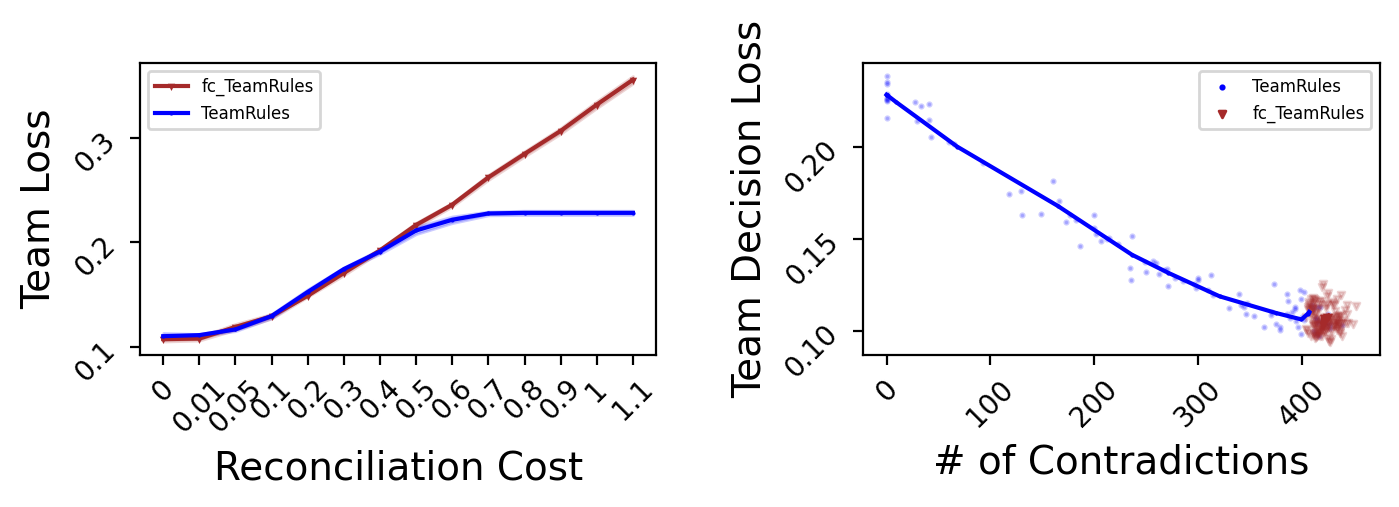}
            \caption[]%
            {{\small Rational}}    
            \label{fig:mean and std of net14}
        \end{subfigure}
        \hfill
        
        \begin{subfigure}[b]{0.725\columnwidth}   
            \centering 
            \includegraphics[width=\textwidth]{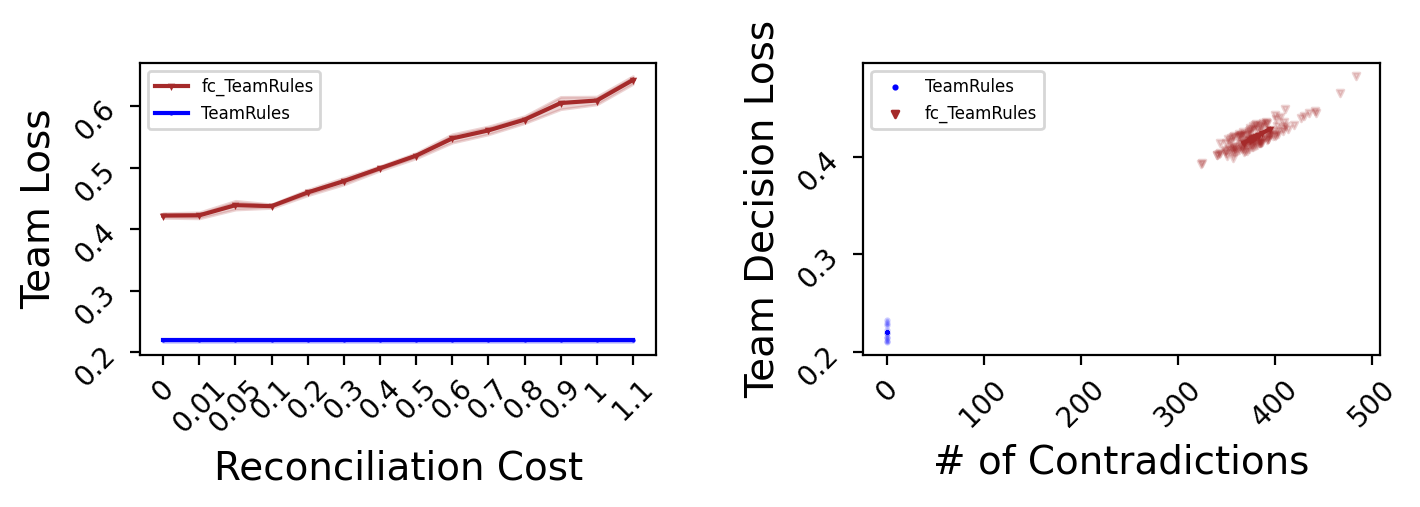}
            \caption[]%
            {{\small Irrational}}    
            \label{fig:mean and std of net44}
        \end{subfigure}
\caption[]{Total team loss for varying reconciliation costs. (Left) Total Team Loss vs. Reconciliation cost. Results show average $+-$ SE. (Right) TDL vs. Number of contradictions. Line connects averages for different cost settings. Transparent points show individual results for cost$+$run.}
\label{fig:selective} 
\end{figure}

Finally, to gauge the value of the selective recommendations mechanism, we create an altered version of \textsc{tr} called \emph{full coverage TeamRules} (\textsc{fc\_tr}). \textsc{fc\_tr} functions just like \textsc{tr}, however, it must always make a recommendation to the human, allowing the comparison between \textsc{tr} and \textsc{fc\_tr} to serve as an ablation study of the selective recommendation mechanism. Again assuming perfect knowledge of the DM's ADB, we observe the performance in terms of TTL and TDL of \textsc{tr} and \textsc{fc\_tr} as reconciliation costs vary in Figure \ref{fig:selective}. In the figures on the left, we see \textsc{tr} leads to an equivalent or lower team loss than \textsc{fc\_tr} for all reconciliation costs. In all cases, because \textsc{fc\_tr} is unable to withhold advice, it is also unable to adjust to rising reconciliation costs. Additionally, even with low reconciliation costs, in some settings such as the Neutral and Irrational settings in the FICO dataset, \textsc{tr} is able to achieve lower TDL by focusing on correcting only the instances that would be beneficial to team performance and allowing the human to focus on the rest. \textsc{fc\_tr} provides recommendations even if they are likely to convince the DM to make incorrect decisions, leading to a high number of contradictions and lower TDL simultaneously. 

In the plots on the right of Figure \ref{fig:selective}, we observe that \textsc{tr} leads to equivalent or lower TDL while making fewer contradictions than \textsc{fc\_tr}. In the rational case with reconciliation costs set to 0, the DM decisions provide little benefit to \textsc{tr} and \textsc{fc\_tr} while contradictions do not harm team performance. In this case, the best course of action is for \textsc{tr} to cover whichever instances it needs to improve on team decision accuracy. Consequently, forcing \textsc{tr} to always make recommendations is not harmful, and \textsc{tr} and \textsc{fc\_tr} lead to the same result. As reconciliation costs rise, \textsc{tr} adjusts and makes fewer contradictions at the cost of TDL. In the irrational and neutral settings, the human has expertise in the accept region, so \textsc{tr} is able to rely on this expertise and only needs to identify rules which improve on the human's area of weakness in their accept region, while \textsc{fc\_tr}, by always having to provide recommendations, must give advice to the human even when the human is superior in their accept region, ultimately harming TDL. 

\section{Conclusion}
In this work, we identify key properties of AI that benefit AI systems' ability to learn to advise humans in high-stakes AIaDM settings. We then introduce the AIaT-Learning framework and instantiate it to develop the \textsc{tr} method which selectively provides inherently interpretable rule-based recommendations by leveraging the human partner's ADB and accounting for the partner's decision history and tolerance for reconciling contradicting recommendations. We show that our framework and method matches or leads to significantly improved team performance over relevant alternatives for various settings defined by a decision task, simulated human behavior, and incurred reconciliation costs. AIaDM settings are common in practice, and our study shows that \emph{methods that advise humans can improve team performance if they include mechanisms that leverage the individual human partner's unique expertise and behavioral tendencies.} While this conclusion is supported by our extensive set of experiments in simulated settings, we hope that our work is an exciting first step that will inspire new research towards the development of advising models and that it will open up research opportunities across disciplines.

\paragraph{Future Work}  Researchers may consider conducting experiments with expert decision-makers in various domains to better gauge opportunity areas for and feasibility of advising methods across contexts. Such experiments would require \emph{extended engagements} with expert DMs to learn of their decision and discretion behaviors followed by the development of a personalized AI advisor for each DM. Certain domains may contain additional AIaDM properties which we did not consider in this work. Additionally, domain-inspired challenges may require computational adjustments to the advising model proposed in this work. For example, it may be necessary to change the optimization scheme or include additional mechanisms as part of the model. Further, researchers can focus on the behavioral challenges that arise from personalized advising methods. Crucially in this work, we build on prior work that demonstrated that the human's discretion to accept recommendations is a function of the human's confidence in their own decisions. However, there is room for future work to develop superior methods to model and predict the human's discretion towards the AI's advice, including specifically to benefit learning-to-advise methods which we demonstrate can be meaningfully informed by such estimations. 
Future work can also develop superior and more efficient methods of interacting with decision-makers to model their discretion behavior, so as to reduce the time and effort needed to adequately model and thereby complement the human decision-maker.

\bibliographystyle{named}
\bibliography{ijcai23}

\begin{thebibliography}{}

\bibitem[\protect\citeauthoryear{Angelino \bgroup \em et al.\egroup
  }{2018}]{angelino2018learning}
Elaine Angelino, Nicholas Larus-Stone, Daniel Alabi, Margo Seltzer, and Cynthia
  Rudin.
\newblock Learning certifiably optimal rule lists for categorical data.
\newblock {\em Journal of Machine Learning Research}, 18:1--78, 2018.

\bibitem[\protect\citeauthoryear{Balagopal \bgroup \em et al.\egroup
  }{2021}]{BALAGOPAL2021102101}
Anjali Balagopal, Dan Nguyen, Howard Morgan, Yaochung Weng, Michael Dohopolski,
  Mu-Han Lin, Azar~Sadeghnejad Barkousaraie, Yesenia Gonzalez, Aurelie Garant,
  Neil Desai, Raquibul Hannan, and Steve Jiang.
\newblock A deep learning-based framework for segmenting invisible clinical
  target volumes with estimated uncertainties for post-operative prostate
  cancer radiotherapy.
\newblock {\em Medical Image Analysis}, 72:102101, 2021.

\bibitem[\protect\citeauthoryear{Bansal \bgroup \em et al.\egroup
  }{2021a}]{bansal_is_2021}
Gagan Bansal, Besmira Nushi, Ece Kamar, Eric Horvitz, and Daniel~S. Weld.
\newblock Is the most accurate ai the best teammate? optimizing ai for
  teamwork.
\newblock {\em Proceedings of the AAAI Conference on Artificial Intelligence},
  35(13):11405--11414, May 2021.

\bibitem[\protect\citeauthoryear{Bansal \bgroup \em et al.\egroup
  }{2021b}]{bansal_does_2021}
Gagan Bansal, Tongshuang Wu, Joyce Zhou, Raymond Fok, Besmira Nushi, Ece Kamar,
  Marco~Tulio Ribeiro, and Daniel Weld.
\newblock Does the whole exceed its parts? the effect of ai explanations on
  complementary team performance.
\newblock In {\em Proceedings of the 2021 CHI Conference on Human Factors in
  Computing Systems}, CHI '21, New York, NY, USA, 2021. Association for
  Computing Machinery.

\bibitem[\protect\citeauthoryear{Basu \bgroup \em et al.\egroup
  }{2021}]{basu_human_2021}
Saunak Basu, Aravinda Garimella, Wencui Han, and Alan Dennis.
\newblock Human {Decision} {Making} in {AI} {Augmented} {Systems}: {Evidence}
  from the {Initial} {Coin} {Offering} {Market}.
\newblock In {\em Proceedings of the 2021 Hawaii International Conference on
  System Sciences}, 2021.

\bibitem[\protect\citeauthoryear{Bertsimas and Tsitsiklis}{1993}]{sim_anneal}
Dimitris Bertsimas and John Tsitsiklis.
\newblock {Simulated Annealing}.
\newblock {\em Statistical Science}, 8(1):10 -- 15, 1993.

\bibitem[\protect\citeauthoryear{Bondi \bgroup \em et al.\egroup
  }{2022}]{Bondi_Koster_Sheahan_Chadwick_Bachrach_Cemgil_Paquet_Dvijotham_2022}
Elizabeth Bondi, Raphael Koster, Hannah Sheahan, Martin Chadwick, Yoram
  Bachrach, Taylan Cemgil, Ulrich Paquet, and Krishnamurthy Dvijotham.
\newblock Role of human-ai interaction in selective prediction.
\newblock {\em Proceedings of the AAAI Conference on Artificial Intelligence},
  36(5):5286--5294, Jun. 2022.

\bibitem[\protect\citeauthoryear{Cai \bgroup \em et al.\egroup
  }{2019}]{cai_human-centered_2019}
Carrie~J. Cai, Emily Reif, Narayan Hegde, Jason Hipp, Been Kim, Daniel Smilkov,
  Martin Wattenberg, Fernanda Viegas, Greg~S. Corrado, Martin~C. Stumpe, and
  Michael Terry.
\newblock Human-{Centered} {Tools} for {Coping} with {Imperfect} {Algorithms}
  {During} {Medical} {Decision}-{Making}.
\newblock In {\em Proceedings of the 2019 {CHI} {Conference} on {Human}
  {Factors} in {Computing} {Systems}}, pages 1--14, Glasgow Scotland Uk, May
  2019. ACM.

\bibitem[\protect\citeauthoryear{Caruana \bgroup \em et al.\egroup
  }{2015}]{rich}
Rich Caruana, Yin Lou, Johannes Gehrke, Paul Koch, Marc Sturm, and Noemie
  Elhadad.
\newblock Intelligible models for healthcare: Predicting pneumonia risk and
  hospital 30-day readmission.
\newblock In {\em Proceedings of the 21th ACM SIGKDD International Conference
  on Knowledge Discovery and Data Mining}, KDD '15, page 1721–1730, New York,
  NY, USA, 2015. Association for Computing Machinery.

\bibitem[\protect\citeauthoryear{Castelo \bgroup \em et al.\egroup
  }{2019}]{task-dependent-algorithm-aversion}
Noah Castelo, Maarten~W. Bos, and Donald~R. Lehmann.
\newblock Task-dependent algorithm aversion.
\newblock {\em Journal of Marketing Research}, 56(5):809--825, 2019.

\bibitem[\protect\citeauthoryear{Chen and Guestrin}{2016}]{xgboost}
Tianqi Chen and Carlos Guestrin.
\newblock Xgboost: A scalable tree boosting system.
\newblock In {\em Proceedings of the 22nd ACM SIGKDD International Conference
  on Knowledge Discovery and Data Mining}, KDD '16, page 785–794, New York,
  NY, USA, 2016. Association for Computing Machinery.

\bibitem[\protect\citeauthoryear{Chiang and Yin}{2021}]{stop}
Chun-Wei Chiang and Ming Yin.
\newblock You’d better stop! understanding human reliance on machine learning
  models under covariate shift.
\newblock In {\em 13th ACM Web Science Conference 2021}, WebSci '21, page
  120–129, New York, NY, USA, 2021. Association for Computing Machinery.

\bibitem[\protect\citeauthoryear{Chong \bgroup \em et al.\egroup
  }{2022}]{CHONG2022107018}
Leah Chong, Guanglu Zhang, Kosa Goucher-Lambert, Kenneth Kotovsky, and Jonathan
  Cagan.
\newblock Human confidence in artificial intelligence and in themselves: The
  evolution and impact of confidence on adoption of ai advice.
\newblock {\em Computers in Human Behavior}, 127:107018, 2022.

\bibitem[\protect\citeauthoryear{Dietvorst \bgroup \em et al.\egroup
  }{2015}]{dietvorst_algorithm_2015}
Berkeley~J. Dietvorst, Joseph~P. Simmons, and Cade Massey.
\newblock Algorithm aversion: {People} erroneously avoid algorithms after
  seeing them err.
\newblock {\em Journal of Experimental Psychology: General}, 144(1):114--126,
  February 2015.
\newblock Publisher: American Psychological Association.

\bibitem[\protect\citeauthoryear{Doshi-Velez and
  Kim}{2017}]{doshi-velez_towards_2017}
Finale Doshi-Velez and Been Kim.
\newblock Towards {A} {Rigorous} {Science} of {Interpretable} {Machine}
  {Learning}.
\newblock {\em arXiv:1702.08608 [cs, stat]}, March 2017.
\newblock arXiv: 1702.08608.

\bibitem[\protect\citeauthoryear{Dua and Graff}{2017}]{uci_adult}
Dheeru Dua and Casey Graff.
\newblock {UCI} machine learning repository, 2017.

\bibitem[\protect\citeauthoryear{FICO}{2018}]{fico_2018}
FICO.
\newblock Fico explainable machine learning chal- lenge., 2018.

\bibitem[\protect\citeauthoryear{Gao \bgroup \em et al.\egroup
  }{2021}]{gao_2021}
Ruijiang Gao, Maytal Saar-Tsechansky, Maria De-Arteaga, Ligong Han, Min~Kyung
  Lee, and Matthew Lease.
\newblock Human-ai collaboration with bandit feedback.
\newblock In Zhi-Hua Zhou, editor, {\em Proceedings of the Thirtieth
  International Joint Conference on Artificial Intelligence, {IJCAI-21}}, pages
  1722--1728. International Joint Conferences on Artificial Intelligence
  Organization, 8 2021.
\newblock Main Track.

\bibitem[\protect\citeauthoryear{Goodman and
  Flaxman}{2017}]{goodman_european_2017}
Bryce Goodman and Seth Flaxman.
\newblock European union regulations on algorithmic decision-making and a
  “right to explanation”.
\newblock {\em AI Magazine}, 38(3):50--57, Oct. 2017.

\bibitem[\protect\citeauthoryear{Green and Chen}{2019a}]{green_disparate_2019}
Ben Green and Yiling Chen.
\newblock Disparate {Interactions}: {An} {Algorithm}-in-the-{Loop} {Analysis}
  of {Fairness} in {Risk} {Assessments}.
\newblock In {\em Proceedings of the {Conference} on {Fairness},
  {Accountability}, and {Transparency}}, pages 90--99, Atlanta GA USA, January
  2019. ACM.

\bibitem[\protect\citeauthoryear{Green and Chen}{2019b}]{green_disparate}
Ben Green and Yiling Chen.
\newblock Disparate interactions: An algorithm-in-the-loop analysis of fairness
  in risk assessments.
\newblock In {\em Proceedings of the Conference on Fairness, Accountability,
  and Transparency}, FAT* '19, page 90–99, New York, NY, USA, 2019.
  Association for Computing Machinery.

\bibitem[\protect\citeauthoryear{Green and Chen}{2019c}]{green_principles}
Ben Green and Yiling Chen.
\newblock The principles and limits of algorithm-in-the-loop decision making.
\newblock {\em Proc. ACM Hum.-Comput. Interact.}, 3(CSCW), nov 2019.

\bibitem[\protect\citeauthoryear{Green and Chen}{2021}]{green_algorithmic_2021}
Ben Green and Yiling Chen.
\newblock Algorithmic risk assessments can alter human decision-making
  processes in high-stakes government contexts.
\newblock {\em Proceedings of the ACM on Human-Computer Interaction}, 5(CSCW2),
  2021.

\bibitem[\protect\citeauthoryear{Johnson}{2021}]{JOHNSON2021203}
Jeff Johnson.
\newblock Chapter 12 - human decision-making is rarely rational.
\newblock In Jeff Johnson, editor, {\em Designing with the Mind in Mind (Third
  Edition)}, pages 203--223. Morgan Kaufmann, third edition edition, 2021.

\bibitem[\protect\citeauthoryear{Jussupow \bgroup \em et al.\egroup
  }{2020}]{jussupow_why_2020}
Ekaterina Jussupow, Izak Benbasat, and Armin Heinzl.
\newblock {WHY} {ARE} {WE} {AVERSE} {TOWARDS} {ALGORITHMS}? {A} {COMPREHENSIVE}
  {LITERATURE} {REVIEW} {ON} {ALGORITHM} {AVERSION}.
\newblock {\em In Proceedings of the 28th European Conference on Information
  Systems (ECIS), An Online AIS Conference}, June 2020.

\bibitem[\protect\citeauthoryear{Keswani \bgroup \em et al.\egroup
  }{2021}]{defer_multiple_2021}
Vijay Keswani, Matthew Lease, and Krishnaram Kenthapadi.
\newblock {\em Towards Unbiased and Accurate Deferral to Multiple Experts},
  page 154–165.
\newblock Association for Computing Machinery, New York, NY, USA, 2021.

\bibitem[\protect\citeauthoryear{Kim \bgroup \em et al.\egroup
  }{2020}]{kim_when_2020}
Antino Kim, Mochen Yang, and Jingjng Zhang.
\newblock When algorithms err: Differential impact of early vs. late errors on
  users' reliance on algorithms.
\newblock 2020.

\bibitem[\protect\citeauthoryear{Klayman \bgroup \em et al.\egroup
  }{1999}]{KLAYMAN1999216}
Joshua Klayman, Jack~B. Soll, Claudia González-Vallejo, and Sema Barlas.
\newblock Overconfidence: It depends on how, what, and whom you ask.
\newblock {\em Organizational Behavior and Human Decision Processes},
  79(3):216--247, 1999.

\bibitem[\protect\citeauthoryear{Lebovitz \bgroup \em et al.\egroup
  }{2022}]{lebovitz_engage_2022}
Sarah Lebovitz, Hila Lifshitz-Assaf, and Natalia Levina.
\newblock To {Engage} or {Not} to {Engage} with {AI} for {Critical}
  {Judgments}: {How} {Professionals} {Deal} with {Opacity} {When} {Using} {AI}
  for {Medical} {Diagnosis}.
\newblock {\em Organization Science}, page orsc.2021.1549, January 2022.

\bibitem[\protect\citeauthoryear{Letham \bgroup \em et al.\egroup
  }{2015}]{rules_why/15-AOAS848}
Benjamin Letham, Cynthia Rudin, Tyler~H. McCormick, and David Madigan.
\newblock {Interpretable classifiers using rules and Bayesian analysis:
  Building a better stroke prediction model}.
\newblock {\em The Annals of Applied Statistics}, 9(3):1350 -- 1371, 2015.

\bibitem[\protect\citeauthoryear{Logg \bgroup \em et al.\egroup
  }{2019}]{LOGG201990}
Jennifer~M. Logg, Julia~A. Minson, and Don~A. Moore.
\newblock Algorithm appreciation: People prefer algorithmic to human judgment.
\newblock {\em Organizational Behavior and Human Decision Processes},
  151:90--103, 2019.

\bibitem[\protect\citeauthoryear{Madras \bgroup \em et al.\egroup
  }{2018}]{madras_predict_2018}
David Madras, Toni Pitassi, and Richard Zemel.
\newblock Predict {Responsibly}: {Improving} {Fairness} and {Accuracy} by
  {Learning} to {Defer}.
\newblock In {\em Advances in {Neural} {Information} {Processing} {Systems}},
  volume~31. Curran Associates, Inc., 2018.

\bibitem[\protect\citeauthoryear{Mahmud \bgroup \em et al.\egroup
  }{2022}]{MAHMUD2022121390}
Hasan Mahmud, A.K.M.~Najmul Islam, Syed~Ishtiaque Ahmed, and Kari Smolander.
\newblock What influences algorithmic decision-making? a systematic literature
  review on algorithm aversion.
\newblock {\em Technological Forecasting and Social Change}, 175:121390, 2022.

\bibitem[\protect\citeauthoryear{Rudin}{2019}]{rudin2019stop}
Cynthia Rudin.
\newblock Stop explaining black box machine learning models for high stakes
  decisions and use interpretable models instead.
\newblock {\em Nature Machine Intelligence}, 1(5):206--215, 2019.

\bibitem[\protect\citeauthoryear{Soares and
  Angelov}{2019}]{soares_fair-by-design_2019}
Eduardo Soares and Plamen Angelov.
\newblock Fair-by-design explainable models for prediction of recidivism.
\newblock {\em arXiv:1910.02043 [cs, stat]}, September 2019.
\newblock arXiv: 1910.02043.

\bibitem[\protect\citeauthoryear{Vellido}{2020}]{vellido_importance_2020}
Alfredo Vellido.
\newblock The importance of interpretability and visualization in machine
  learning for applications in medicine and health care.
\newblock {\em Neural Computing and Applications}, 32(24):18069--18083,
  December 2020.

\bibitem[\protect\citeauthoryear{Wang and
  Saar-Tsechansky}{2020}]{wang_augmented_2020}
Tong Wang and Maytal Saar-Tsechansky.
\newblock Augmented {Fairness}: {An} {Interpretable} {Model} {Augmenting}
  {Decision}-{Makers}' {Fairness}.
\newblock {\em arXiv:2011.08398 [cs]}, November 2020.
\newblock arXiv: 2011.08398.

\bibitem[\protect\citeauthoryear{Wang \bgroup \em et al.\egroup
  }{2017}]{wang2017bayesian}
Tong Wang, Cynthia Rudin, Finale Doshi-Velez, Yimin Liu, Erica Klampfl, and
  Perry MacNeille.
\newblock A bayesian framework for learning rule sets for interpretable
  classification.
\newblock {\em The Journal of Machine Learning Research}, 18(1):2357--2393,
  2017.

\bibitem[\protect\citeauthoryear{Wang \bgroup \em et al.\egroup
  }{2021}]{wang2021scalable}
Zhuo Wang, Wei Zhang, Ning Liu, and Jianyong Wang.
\newblock Scalable rule-based representation learning for interpretable
  classification.
\newblock {\em Advances in Neural Information Processing Systems},
  34:30479--30491, 2021.

\bibitem[\protect\citeauthoryear{Wang \bgroup \em et al.\egroup
  }{2022a}]{will_you_accept}
Xinru Wang, Zhuoran Lu, and Ming Yin.
\newblock Will you accept the ai recommendation? predicting human behavior in
  ai-assisted decision making.
\newblock In {\em Proceedings of the ACM Web Conference 2022}, WWW '22, page
  1697–1708, New York, NY, USA, 2022. Association for Computing Machinery.

\bibitem[\protect\citeauthoryear{Wang \bgroup \em et al.\egroup
  }{2022b}]{edit_machine}
Zijie~J. Wang, Alex Kale, Harsha Nori, Peter Stella, Mark~E. Nunnally,
  Duen~Horng Chau, Mihaela Vorvoreanu, Jennifer Wortman~Vaughan, and Rich
  Caruana.
\newblock Interpretability, then what? editing machine learning models to
  reflect human knowledge and values.
\newblock In {\em Proceedings of the 28th ACM SIGKDD Conference on Knowledge
  Discovery and Data Mining}, KDD '22, page 4132–4142, New York, NY, USA,
  2022. Association for Computing Machinery.

\bibitem[\protect\citeauthoryear{Wang}{2018}]{wang2018multi}
Tong Wang.
\newblock Multi-value rule sets for interpretable classification with
  feature-efficient representations.
\newblock {\em Advances in neural information processing systems}, 31, 2018.

\bibitem[\protect\citeauthoryear{Wang}{2019}]{wang_gaining_nodate}
Tong Wang.
\newblock Gaining free or low-cost interpretability with interpretable partial
  substitute.
\newblock In Kamalika Chaudhuri and Ruslan Salakhutdinov, editors, {\em
  Proceedings of the 36th International Conference on Machine Learning},
  volume~97 of {\em Proceedings of Machine Learning Research}, pages
  6505--6514. PMLR, 09--15 Jun 2019.

\bibitem[\protect\citeauthoryear{Wilder \bgroup \em et al.\egroup
  }{2021}]{wilder_learning_2020}
Bryan Wilder, Eric Horvitz, and Ece Kamar.
\newblock Learning to complement humans.
\newblock In {\em Proceedings of the Twenty-Ninth International Joint
  Conference on Artificial Intelligence}, IJCAI'20, 2021.

\bibitem[\protect\citeauthoryear{Yildiz}{2014}]{vc_rules}
Olcay~Taner Yildiz.
\newblock Vc-dimension of rule sets.
\newblock In {\em 2014 22nd International Conference on Pattern Recognition},
  pages 3576--3581, 2014.

\bibitem[\protect\citeauthoryear{Zhang \bgroup \em et al.\egroup
  }{2022}]{complete-me}
Qiaoning Zhang, Matthew~L Lee, and Scott Carter.
\newblock You complete me: Human-ai teams and complementary expertise.
\newblock In {\em Proceedings of the 2022 CHI Conference on Human Factors in
  Computing Systems}, CHI '22, New York, NY, USA, 2022. Association for
  Computing Machinery.

\end{thebibliography}

\clearpage

\appendix
\section{Objective} \label{app:objective}
The \textsc{tr} objective is set to the team loss function:
 \begin{equation*} \label{eq:utility1}
 \textstyle
     \mathcal{L}(D,R) = \ell(D,R) + \omega(D,R)
 \end{equation*}

 \noindent Where $\ell(D,R)$ is the \emph{discretion adjusted} decision loss: 
 \begin{equation*}
 \begin{aligned}
 \tiny
     \ell(D, R) &= \sum_{i=1}^n  \bigg[p(a_i)\bigg(\big((1-y_i)\mathbbm{C}(R^+, x_i)\big) \\ &+ \big(y_i(1-\mathbbm{C}(R^+, x_i))\big)\mathbbm{C}(R^-, x_i)  \\ 
     &+ (1-\mathbbm{C}(R^+, x_i))(1-\mathbbm{C}(R^-, x_i)) \\&\times (y_i(1-h_i) + h_i(1-y_i)) \bigg)\bigg]
 \end{aligned}
 \normalsize
 \end{equation*}

 \noindent And $\omega(D,R)$ represents the costs of reconciling contradicting AI recommendations incurred by the human DM:
 \begin{equation*}
     \begin{aligned}
 \small
 \textstyle
     \omega(D, R) &= \alpha\sum_{i=1}^n  \bigg[\bigg(\big((1-h_i)\mathbbm{C}(R^+, x_i)\big) \\&+ \big(h_i(1-\mathbbm{C}(R^+, x_i))\big)\mathbbm{C}(R^-, x_i) \bigg)\bigg]
 \end{aligned}
 \normalsize
 \end{equation*}

\section{Experiment Details} \label{exp_deets}

\subsection{Checkers Dataset} 
\begin{itemize}
    \item Train Instances: 4,000
    \item Test Instances: 800
    \item Features $x^{[1]}$, $x^{[2]}$ are each i.i.d ~$Uniform(0,2)$ r.v.
    \item $y = \mathbbm{1}\{x^{[1]} \leq 1\} \mathbbm{1}\{x^{[2]} \geq 1\}+ \mathbbm{1}\{x^{[1]} \geq 1\} \mathbbm{1}\{x^{[2]} \leq 1\}$
    \item Human has 80\% accuracy when $x^{[1]} > x^{[2]}$, otherwise has 100\% accuracy. 
    \item Rational Behavior: $a = 1$ (accept) in lower accuracy region, otherwise $a=0$ (reject)
    \item Neutral Behavior: $a = 1$ (accept) when $x^{[1]} >= 1$, otherwise $a=0$ (reject)
    \item Irrational Behavior: $a = 0$ (reject) lower accuracy region, otherwise $a=1$ (accept)
\end{itemize}

\subsection{Gaussian Dataset} 
\begin{outline}
 \1 Train Instances: 4,000
 \1 Test Instances: 800
 \1 Features $x^{[1]}$...$x^{[20]}$ are each i.i.d ~$\mathcal{N}(0,1)$ r.v.
 \1 Given: 
 \2 $\varphi(x)$ is the standard normal p.d.f
 \2 $q_{0.5}(v)$ is the median value of vector v
 \2 $v_1 = \varphi(\sum_{d=1}^2 x^{[d]})$
 \2 $$v_2 = \varphi(\sum_{d=1}^4 x^{[d]}) + \varphi(\sum_{d=5}^8 x^{[d]}) + \varphi(\sum_{d=9}^{16} x^{[d]}) + \varphi(\sum_{d=17}^{20} x^{[d]})$$
 \1 y = 1 IFF: 
 \begin{align*}
        &\bigg[(\sum_{d=1}^{20} x^{[d]} < 0) \text{ AND } \big(v_1) > q_{0.5}(v_1)\big)\bigg] \\
        &\text{OR} \\
        &\bigg[(\sum_{d=1}^{20} x^{[d]} \geq 0) \text{ AND } \big(v_2) < q_{0.5}(v_2)\big)\bigg]
    \end{align*}
Intuitively, if the sum of features is less than 0, the label is a simple function of $x^{[1]}$ and $x^{[2]}$. Otherwise, it is a complex function of all features. 
\1 Human is created by first training a logistic regression model $p(y|x)$ on an excluded subset of the training data. Next, human accuracy is set to 50\% wherever $2|p(y|x) - 0.5| > 0.5$, otherwise human accuracy is set to 100\%. 
\1 Rational Behavior: $a = 1$ (accept) in lower accuracy region, otherwise $a=0$ (reject)
\1  Neutral Behavior: $a = 0$ (accept) when $\sum_{d=1}^{20} x^{[d]} < 0$, otherwise $a = 1$ (reject)
\1 Irrational Behavior: $a = 0$ (reject) lower accuracy region, otherwise $a=1$ (accept)
\end{outline}
\subsection{FICO, Adult, HR Datasets}
\begin{outline}
 \1 Train Instances: 9,459(FICO); 23,499(Adult); 390(HR)
 \1 Test Instances: 1,892(FICO); 4,700(Adult);  78(HR)

 \1 Human is created by first training a logistic regression model $p(y|x)$ on an excluded subset of the training data. Next, human accuracy is set to 50\% wherever: 
 \2 $2|p(y|x) - 0.5| > 0.5$, otherwise human accuracy is set to 100\%. (FICO \& HR)
 \2 $0.5 < 2|p(y|x) - 0.5| < 0.8$. Human accuracy set to 100\% when $0 < 2|p(y|x) - 0.5| < 0.4$. Remaining human decision are original logistic regression model output decision. (Adult)
 \1 Rational Behavior: $a = 1$ (accept) in lower accuracy region, otherwise $a = 0$ (reject)
 \1 Neutral Behavior: $a = 1$ (accept) IFF: 
 \2 $ExternalRiskEstimate65.0 \geq 65$ \\AND  $NumSatisfactoryTrades24.0 \geq 24$. (FICO)
 \2 $occupation\_Exec\text{-}managerial = 1$ (Adult)
 \2 $'RelationshipSatisfaction3.0' \geq 3$ \\AND $StockOptionLevel = 0$. (HR)
 \1 Irrational Behavior: $a = 0$ (reject) lower accuracy region, otherwise $a=1$ (accept)
\end{outline}

\subsection{Training Parameters}
\begin{itemize}
    \item $T = 500$
    \item $C_0 = 0.01$
    \item $\beta_0 = 0.05$
    \item $\beta_1 = 1$ for Checkers dataset, $\beta_1 = 3$ for all other datasets
    \item $\beta_2 = 10,000$
    \item $q = 0.05$
    \item $\tau = 0.5$
    
\end{itemize}

\section{Complete Experiment Results}
\label{all_results}

\begin{figure*}[]
        \centering
        \begin{subfigure}[b]{0.70\columnwidth}
            \centering
            \includegraphics[width=0.70\columnwidth]{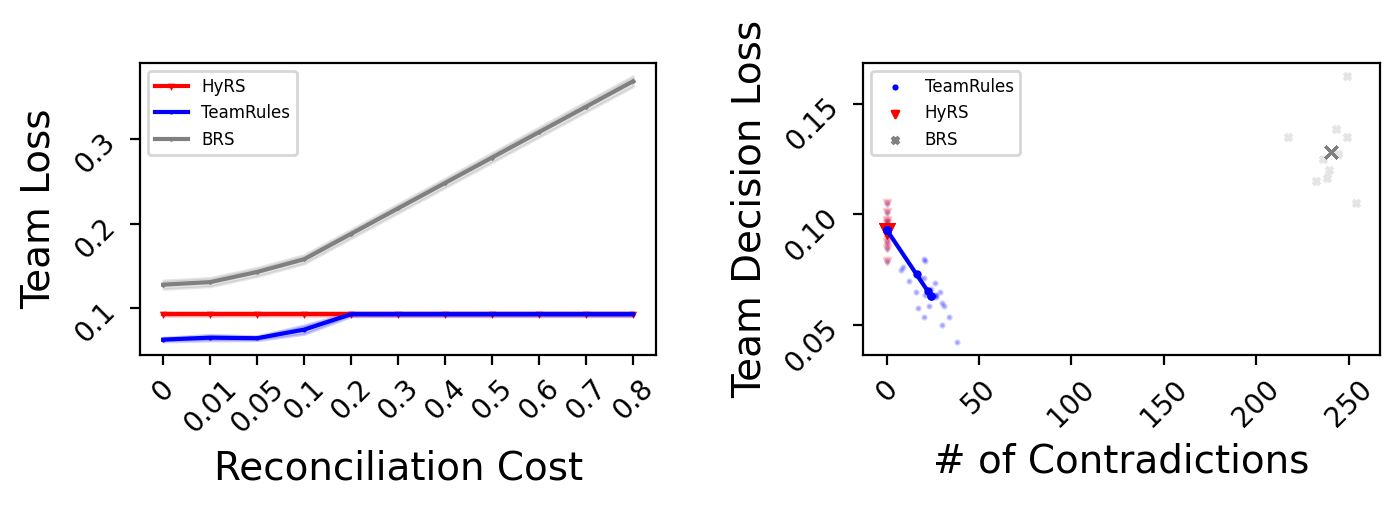}
            \caption[Network2]%
            {{\small Checkers: Rational}}    
        \end{subfigure}
        \hfill
        \begin{subfigure}[b]{0.70\columnwidth}  
            \centering 
            \includegraphics[width=0.70\columnwidth]{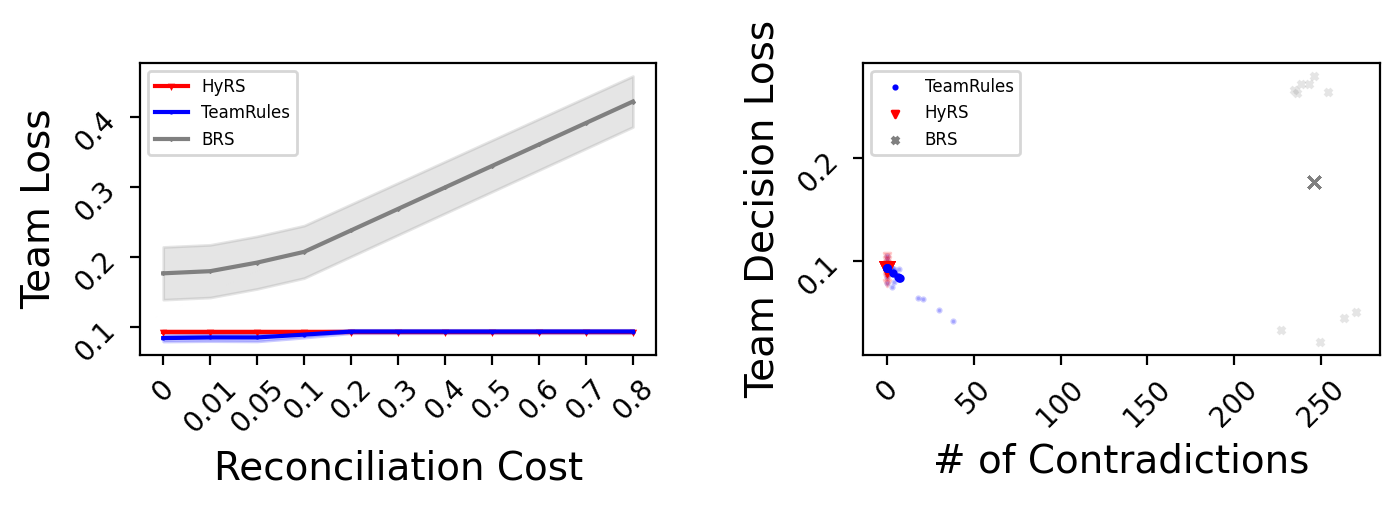}
            \caption[]%
            {{\small Checkers: Neutral}}    
        \end{subfigure}
        \hfill
        \begin{subfigure}[b]{0.70\columnwidth}   
            \centering 
            \includegraphics[width=0.70\columnwidth]{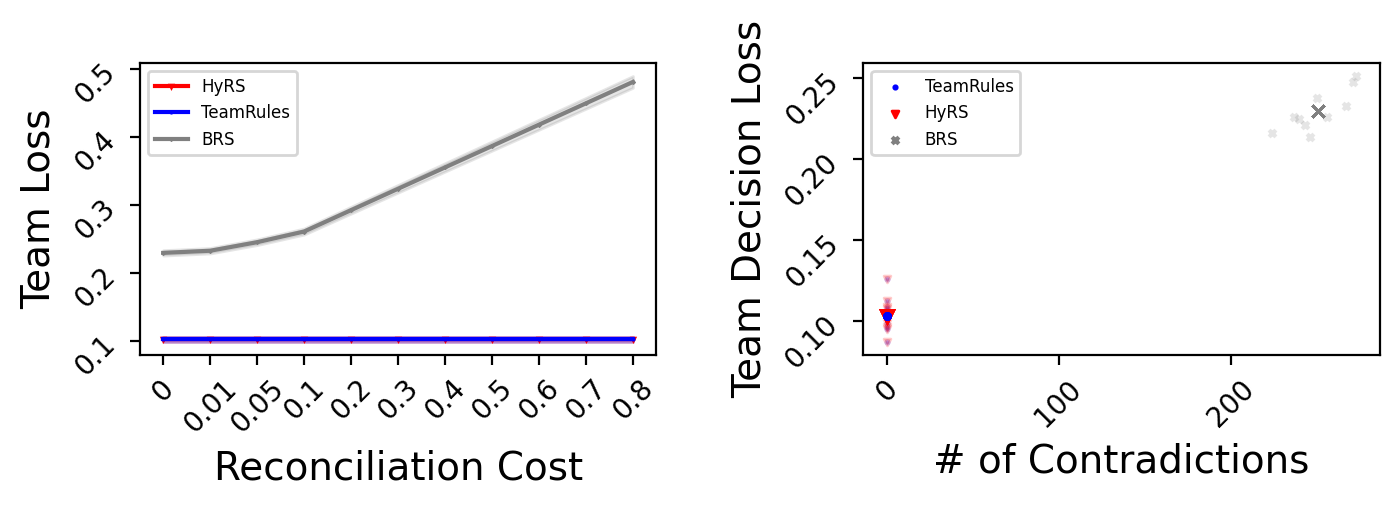}
            \caption[]%
            {{\small Checkers: Irrational}}  
            \label{fig:mean and std of net44}
        \end{subfigure}
        \hfill
        \begin{subfigure}[b]{0.70\columnwidth}
            \centering
            \includegraphics[width=0.70\columnwidth]{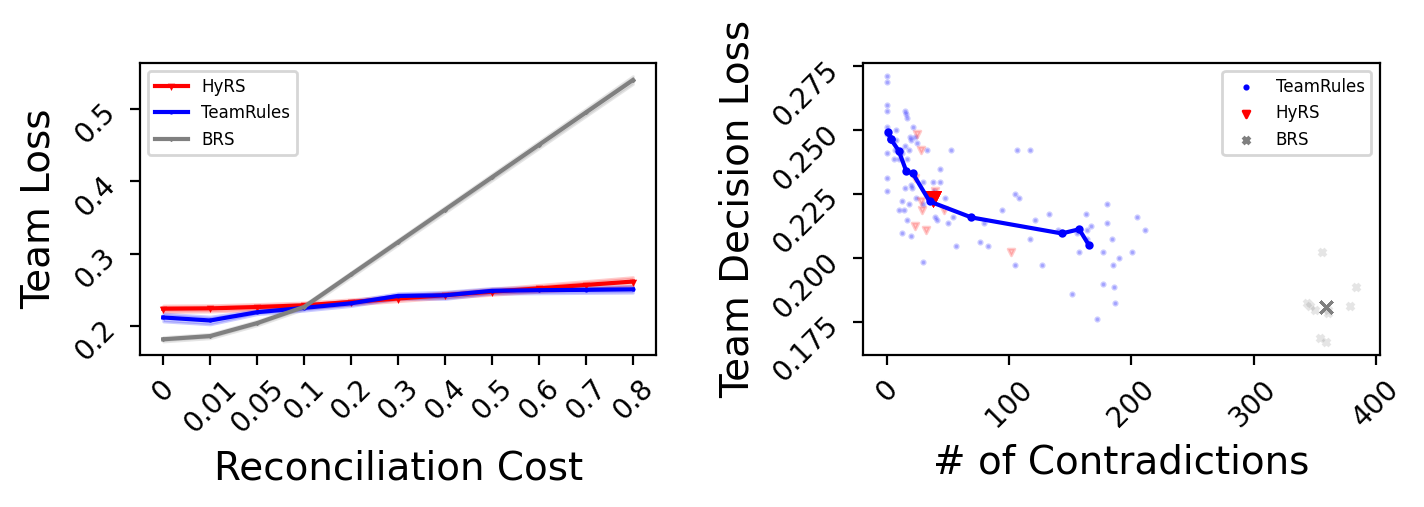}
            \caption[Network2]%
            {{\small Gaussian: Rational}}    
        \end{subfigure}
        \hfill
        \begin{subfigure}[b]{0.70\columnwidth}  
            \centering 
            \includegraphics[width=0.70\columnwidth]{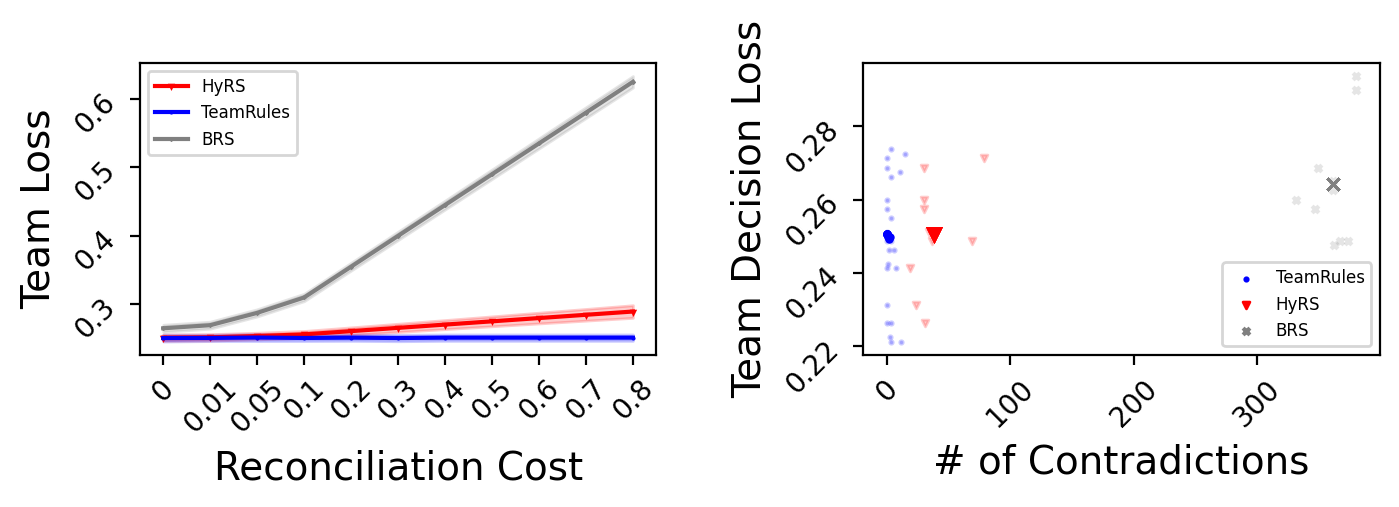}
            \caption[]%
            {{\small Gaussian: Neutral}}    
        \end{subfigure}
        \hfill
        \begin{subfigure}[b]{0.70\columnwidth}   
            \centering 
            \includegraphics[width=0.70\columnwidth]{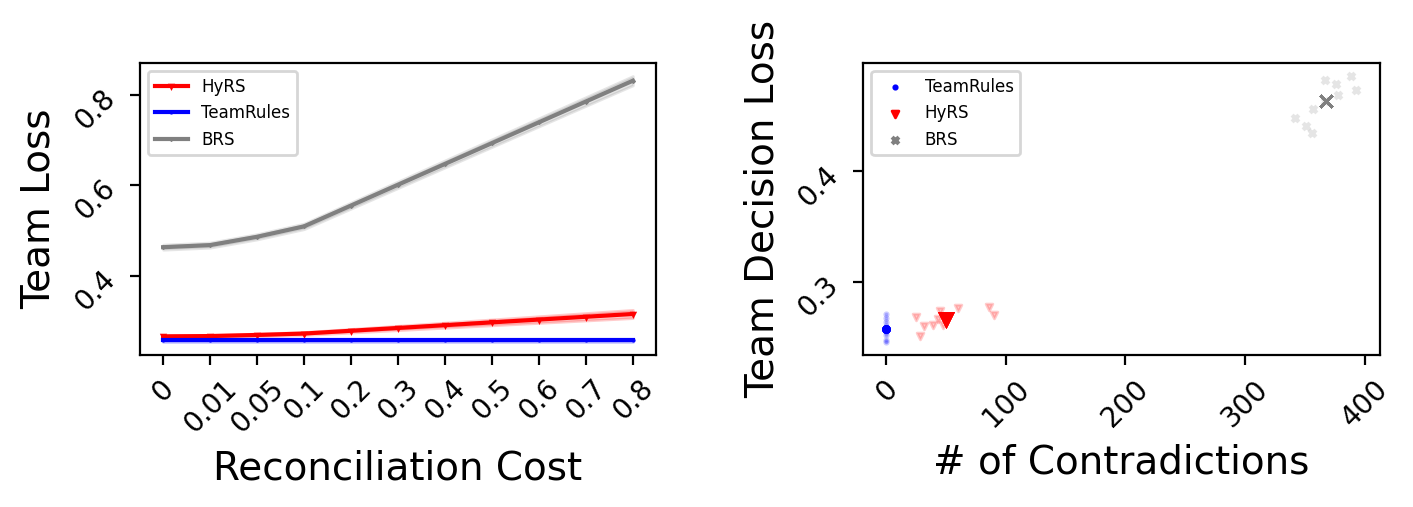}
            \caption[]%
            {{\small Gaussian: Irrational}}  
        \end{subfigure}

\caption[]{Results from varying reconciliation cost parameter for Checkers and Gaussian datasets.}
\label{main_checkers_gauss} 
\end{figure*}

\begin{figure*}[]
        \centering
        \begin{subfigure}[b]{0.70\columnwidth}
            \centering
            \includegraphics[width=0.70\columnwidth]{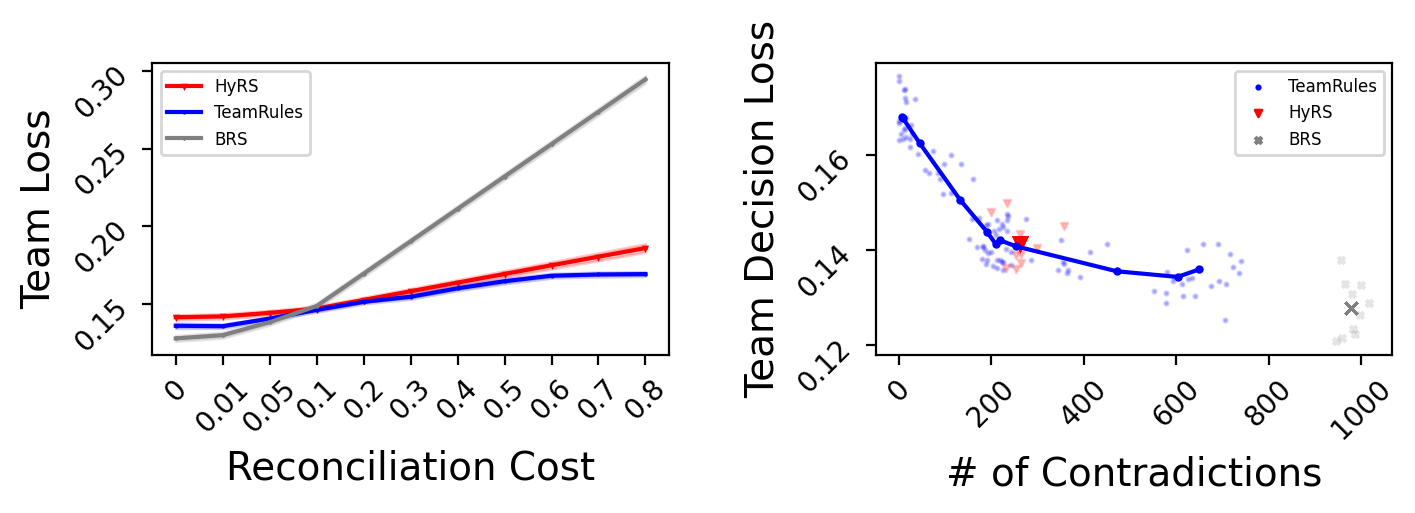}
            \caption[Network2]%
            {{\small Adult: Rational}}    
        \end{subfigure}
        \hfill
        \begin{subfigure}[b]{0.70\columnwidth}  
            \centering 
            \includegraphics[width=0.70\columnwidth]{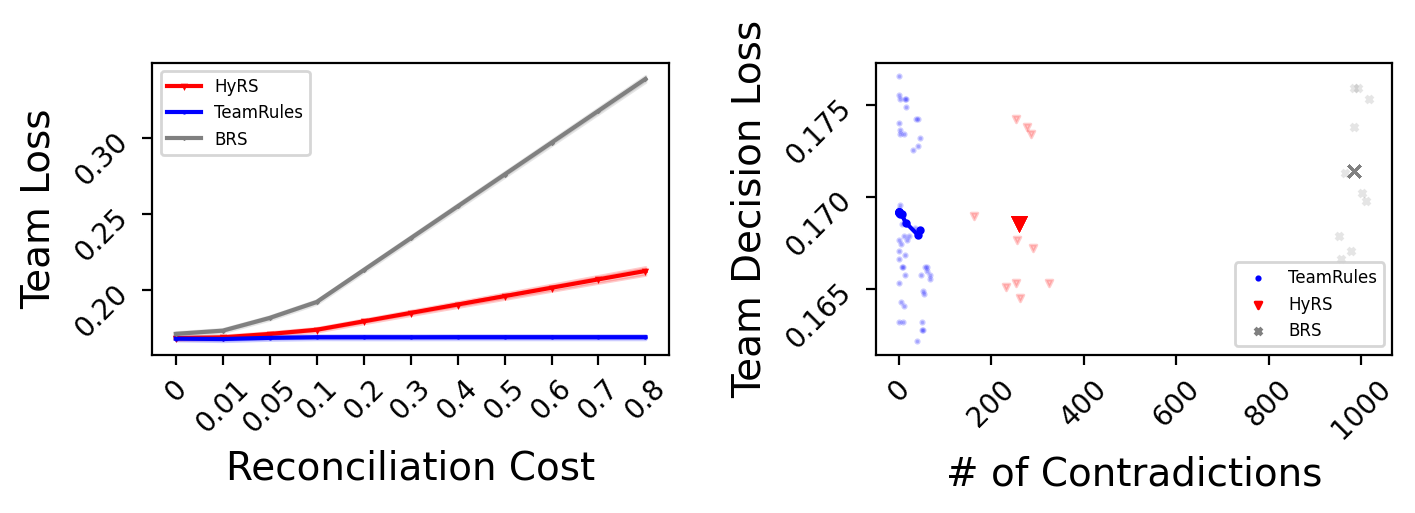}
            \caption[]%
            {{\small Adult: Neutral}}    
        \end{subfigure}
        \hfill
        \begin{subfigure}[b]{0.70\columnwidth}   
            \centering 
            \includegraphics[width=0.70\columnwidth]{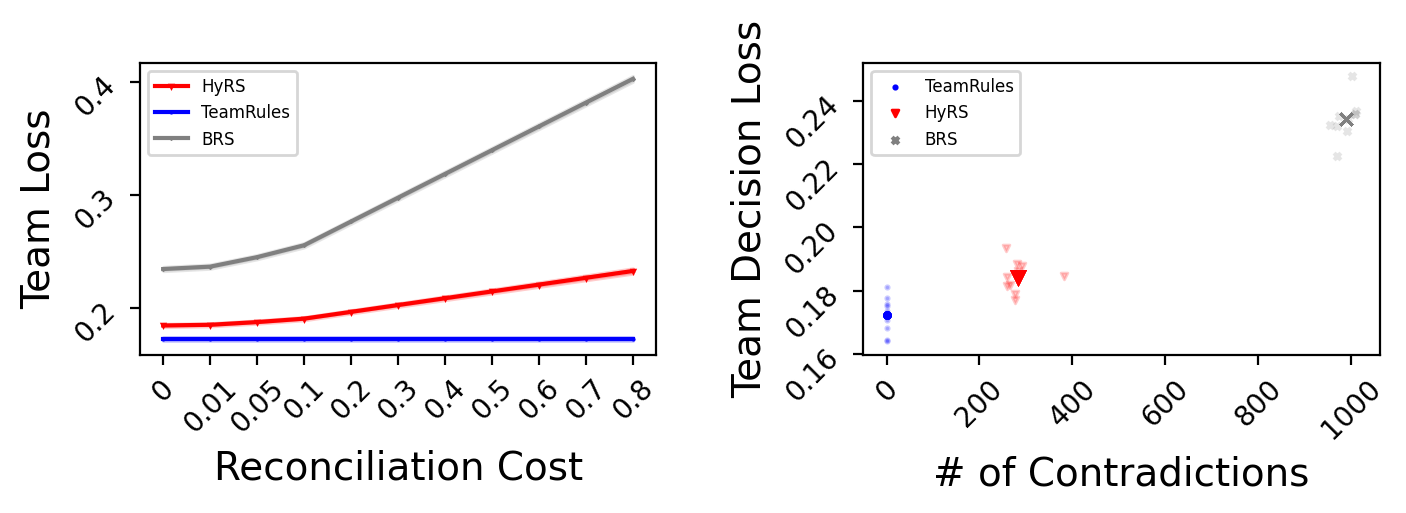}
            \caption[]%
            {{\small Adult: Irrational}}    
        \end{subfigure}
        \hfill
        \begin{subfigure}[b]{0.70\columnwidth}
            \centering
            \includegraphics[width=0.70\columnwidth]{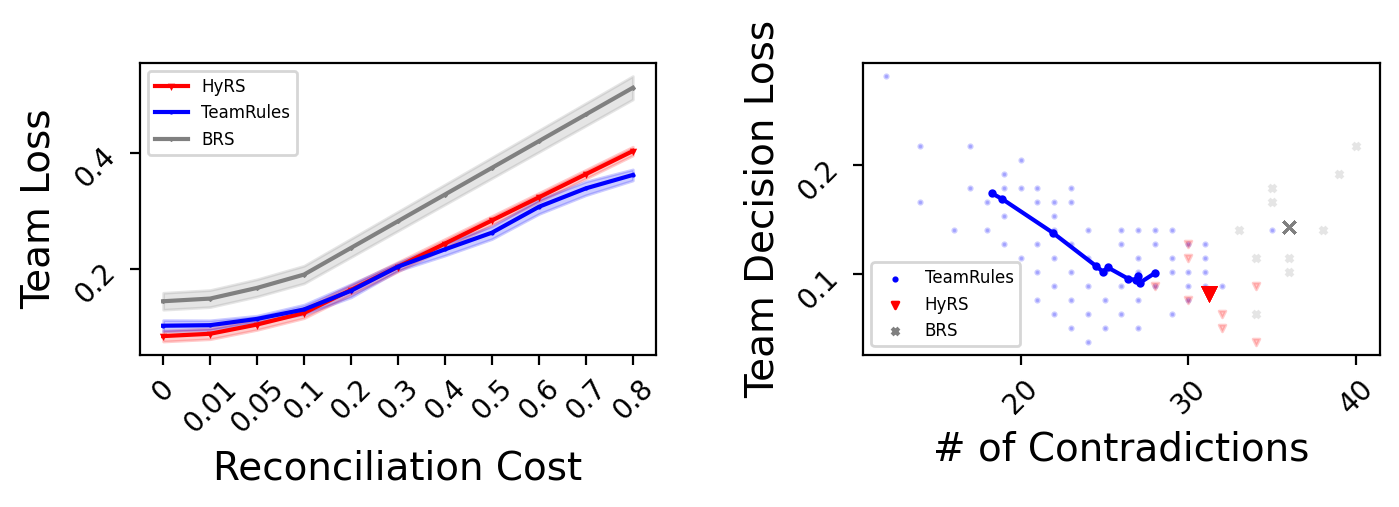}
            \caption[Network2]%
            {{\small HR: Rational}}    
        \end{subfigure}
        \hfill
        \begin{subfigure}[b]{0.70\columnwidth}  
            \centering 
            \includegraphics[width=0.70\columnwidth]{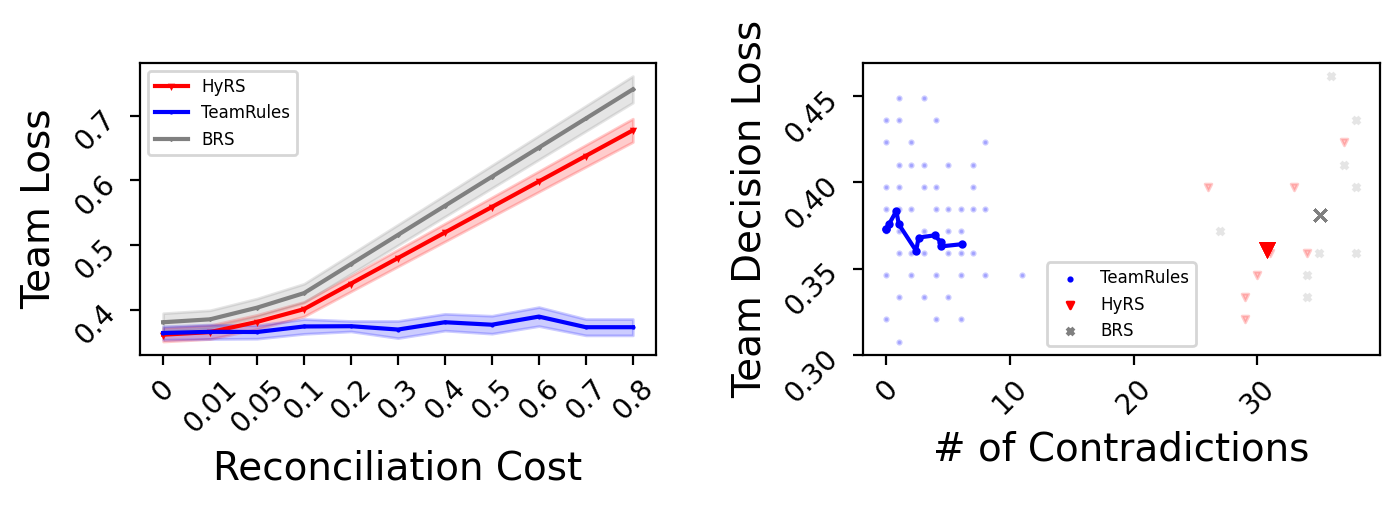}
            \caption[]%
            {{\small HR: Neutral}}    
        \end{subfigure}
        \hfill
        \begin{subfigure}[b]{0.70\columnwidth}   
            \centering 
            \includegraphics[width=0.70\columnwidth]{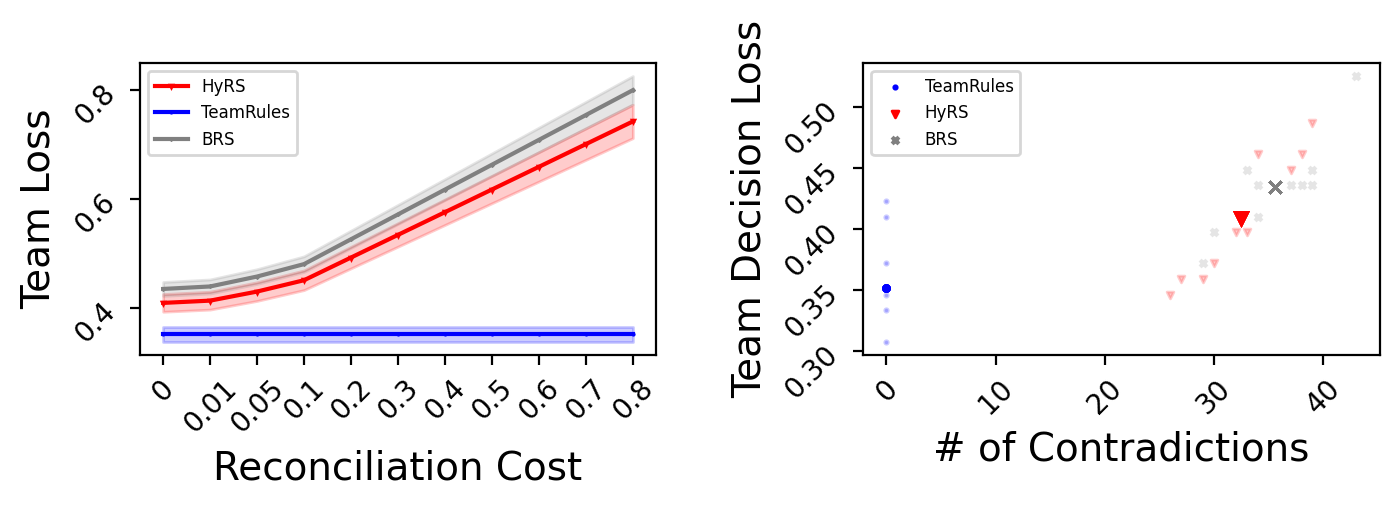}
            \caption[]%
            {{\small HR: Irrational}}    
        \end{subfigure}

\caption[]{Results from varying reconciliation cost parameter for Checkers, Adult and HR datasets.}
\label{main_adult_hr} 
\end{figure*}

\begin{figure}[ht]
    \centering

        \begin{subfigure}[b]{0.3\columnwidth}
            \centering
            \includegraphics[width=\columnwidth]{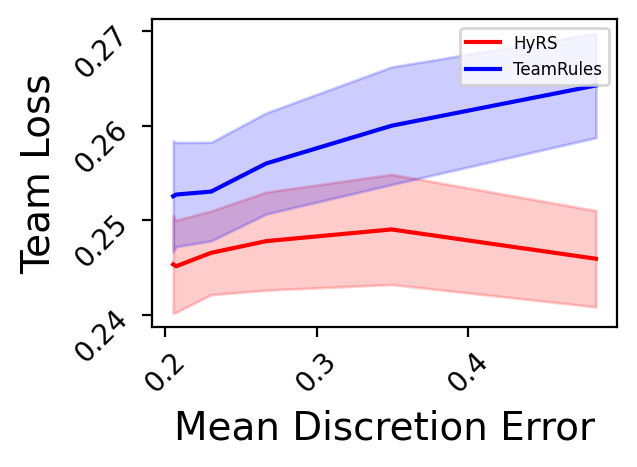}
            \caption[]%
            {{\small Rational}}    
        \end{subfigure}
        \hfill
        \begin{subfigure}[b]{0.3\columnwidth}  
            \centering 
            \includegraphics[width=\columnwidth]{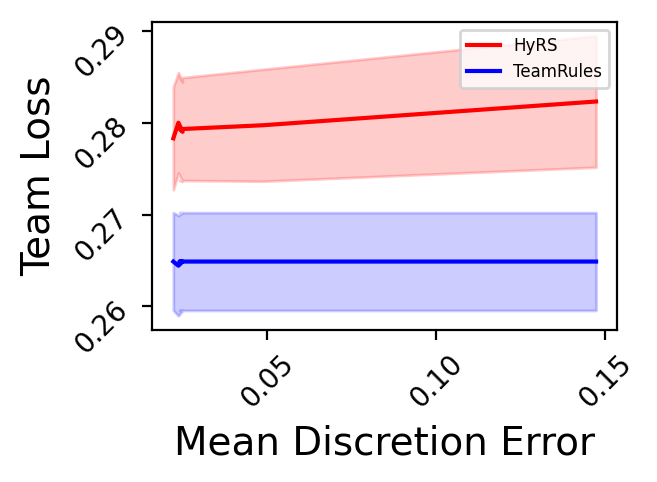}
            \caption[]%
            {{\small Neutral}}    
        \end{subfigure}
        \hfill
        \begin{subfigure}[b]{0.3\columnwidth}   
            \centering 
            \includegraphics[width=\columnwidth]{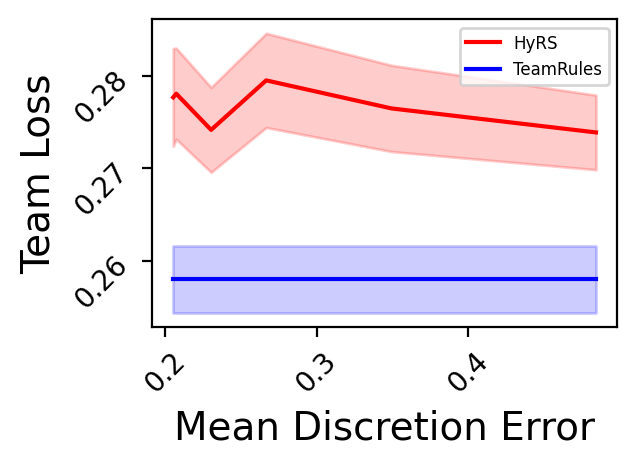}
            \caption[]%
            {{\small Irrational}}    
        \end{subfigure}
    \caption{Total Team Loss on Gaussian data with 0.3 reconciliation cost. Discretion model learned using XGBoost on subsets of data leading to decreasing discretion model accuracy. Avg. of 5 reps. with randomized 80-20\% train-test split on each run.}
    \label{gaussian_err}
\end{figure}

We show all remaining experimental results in Figures \ref{main_checkers_gauss}, \ref{main_adult_hr}, and \ref{gaussian_err}. 

\newpage

\section{Reproducibility}
Code for all methods and experiments will be made publicly available on Github. 

\end{document}